\documentclass[journal,10pt]{IEEEtran}
\usepackage{amsmath}
\usepackage{amssymb}
\usepackage{amsfonts}
\usepackage{graphicx}
\usepackage{epsfig}
\usepackage{subfigure}
\usepackage{psfrag}
\usepackage{cite}
\usepackage{latexsym}
\usepackage{url}
\usepackage{color}

\begin{document}
\title{A General Design Framework for MIMO Wireless Energy Transfer with Limited Feedback
\thanks{J. Xu was with the Department of Electrical and Computer Engineering, National University of Singapore. He is now with the Engineering Systems and Design Pillar, Singapore University of Technology and Design (e-mail:~jiexu.ustc@gmail.com).}
\thanks{R. Zhang is with the Department of Electrical and Computer Engineering, National University of Singapore (e-mail: elezhang@nus.edu.sg). He is also with the Institute for Infocomm Research, A*STAR, Singapore.}}

\author{Jie Xu and Rui Zhang}

\maketitle
\begin{abstract}
Multi-antenna or multiple-input multiple-output (MIMO) technique can significantly improve the efficiency of radio frequency (RF) signal enabled wireless energy transfer (WET). To fully exploit the energy beamforming gain at the energy transmitter (ET), the knowledge of channel state information (CSI) is essential, which, however, is difficult to be obtained in practice due to the energy and hardware limitation of the energy receiver (ER). To overcome this difficulty, under a point-to-point MIMO WET setup, this paper proposes a general design framework for a new type of channel learning method based on the ER's energy measurement feedback. Specifically, the ER measures and encodes the harvested energy levels over different training intervals into bits, and sends them to the ET via a feedback link of limited rate. Based on the energy-level feedback, the ET adjusts transmit beamforming in subsequent training intervals and obtains refined estimates of the MIMO channel by leveraging the technique of analytic center cutting plane method (ACCPM) in convex optimization. Under this general design framework, we further propose two specific feedback schemes based on energy quantization and energy comparison, where the feedback bits at each interval are generated at the ER by quantizing the measured energy level at the current interval and comparing it with those in previous intervals, respectively. Numerical results are provided to compare the performance of the two feedback schemes. It is shown that energy quantization performs better when the number of feedback bits per interval is large, while energy comparison is more effective vice versa.
\end{abstract}

\newtheorem{definition}{\underline{Definition}}[section]
\newtheorem{fact}{Fact}
\newtheorem{assumption}{Assumption}
\newtheorem{theorem}{\underline{Theorem}}[section]
\newtheorem{lemma}{\underline{Lemma}}[section]
\newtheorem{corollary}{\underline{Corollary}}[section]
\newtheorem{proposition}{\underline{Proposition}}[section]
\newtheorem{example}{\underline{Example}}[section]
\newtheorem{remark}{\underline{Remark}}[section]
\newtheorem{algorithm}{\underline{Algorithm}}[section]
\newcommand{\mv}[1]{\mbox{\boldmath{$ #1 $}}}

\section{Introduction}

\begin{figure}
\centering
 \epsfxsize=1\linewidth
    \includegraphics[width=8cm]{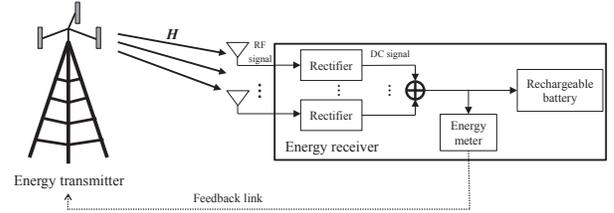}
\caption{A point-to-point MIMO WET system with closed-loop feedback.} \label{fig:System}
\end{figure}

Radio frequency (RF) signal enabled wireless energy transfer (WET) has become a promising technology to provide perpetual, cost-effective, and convenient energy supply to low-power wireless devices such as RF identification (RFID) tags and sensor nodes for various low-power applications in future smart commercial and industrial systems with wireless internet-of-things (IoT) \cite{BiHoZhang2014,BiZengZhang2016}. In RF-based WET systems, dedicated energy transmitters (ETs) are deployed to control and coordinate the transfer of wireless energy via RF signals to a set of distributed energy receivers (ERs), and each ER uses the rectifier at each of its receive antennas (also known as rectennas as a whole) to convert the received RF signals to direct current (DC) signals, which are combined to charge the battery at the ER \cite{ZhouZhangHo2013} (see, e.g., a point-to-point WET system in Fig. \ref{fig:System}). Examples of commercial ERs include the P2110 and P2210 Powerharvester Receivers developed by the Powercast company \cite{Powercast}. Due to the long operating range (say, tens of meters) and flexibility to charge multiple devices at the same time (thanks to the broadcast property of radio signal), RF-based far-field WET has competitive advantages over existing near-field WET techniques such as induction coupling, which generally operates with much shorter distance (say, centimeters) and for point-to-point power transfer only \cite{BiHoZhang2014}.  Furthermore, since RF signal can carry energy as well as modulated information at the same time, a joint investigation of wireless information and energy transfer is appealing, which has recently drawn a great amount of attention. In particular, simultaneous wireless information and power transfer (SWIPT) (see, e.g., \cite{ZhangHo2013,XuLiuZhang2014}) and wireless powered communication network (WPCN) (see, e.g., \cite{JuZhang2014,LiuZhangChua2014}) are two important operation models that have been thoroughly studied in literature. In SWIPT systems, wireless information and energy are transmitted using the same signal; while in WPCN, wireless information transmission is powered  by WET using different signals.

One major practical challenge for implementing RF-based WET is the significant decay of energy transfer efficiency of WET over the distance from the ET to the ER due to the severe prorogation power loss of RF signal. To improve the efficiency, multiple-input multiple-output (MIMO) techniques by equipping multiple antennas at the ET and/or the ER have been proposed as an appealing solution (see, e.g., \cite{ZhangHo2013,XuLiuZhang2014,LiuZhangChua2014,XuBiZhang2015,CheXuDuanZhang2015} and the references therein).{\footnote{In practice, there is a limit on the number of antennas that can be deployed on a device of finite dimensions; however, with the prevalence of exploiting higher frequency bands (e.g., from several to tens of GHz) in future wireless systems, the antenna size will be significantly reduced and as a result it is more practically feasible to install multiple antennas on a single device.}} Multiple antennas at the ET help focus the transmitted wireless energy on the direction to the ER via the technique of digital beamforming or so-called energy beamforming, while multiple antennas at the ER increase the effective aperture area for received RF energy harvesting, both leading to significantly enhanced energy transfer efficiency. This helps extend the operating range of WET with given ERs' energy requirements, and/or enable more power-consuming applications by increasing ERs' harvested energy with fixed distances. However, the benefit of energy beamforming in MIMO WET critically relies on the availability of the channel state information (CSI) at the ET, which needs to be efficiently learned in practical systems.

There are in general three types of channel acquisition methods that have been proposed in the literature for MIMO WET. The first method is by exploiting the channel reciprocity between the forward (from the ET to the ER) and reverse (from the ER to the ET) links \cite{ZengZhang2014A,ZengZhang2014B,YangHoZhangGuan2014}, where the ET obtains the forward link CSI by performing a reverse link channel estimation based on the training signals sent by the ER. It is worth noting that this method is only applicable for time-division duplex (TDD) based systems, where the condition of channel reciprocity practically holds with good accuracy.

The second method is via sending training signals directly in the forward link from the ET to the ER, through which the ER estimates the MIMO channel and then sends the estimated channel back to the ET via the reverse link \cite{YangHoGuan2014}. This method is reminiscent of the conventional channel estimation and feedback approach in wireless communication (see, e.g., \cite{Love2008} and the references therein), which is applicable for both TDD and frequency-division duplex (FDD) based systems. However, under the new setup of WET (instead of wireless communication), this method requires the ER (not an information receiver) to implement additional baseband signal processing for channel estimation, which is not feasible with the existing hardware at the ER designed only for RF energy harvesting (instead of information decoding) as shown in Fig. \ref{fig:System}.

To cater for the hardware limitation of existing ERs, in our previous work \cite{XuZhang_OneBit} we proposed a third channel learning method for MIMO WET based on the energy measurement feedback by the ER. Specifically, the ER measures its harvested energy levels over different training intervals using an energy meter (see Fig. \ref{fig:System}), and sends one feedback bit to the ET per interval to indicate the increase or decrease of the measured energy in the current interval as compared to that in the previous interval. Based on the feedback bits collected in the present and past intervals, the ET adjusts its transmit beamforming in subsequent training intervals and obtains refined estimates of the MIMO channel. Note that similar channel learning methods with one-bit feedback have also been studied in cognitive radio MIMO systems \cite{NoamGoldsmith2013,GopalakrishnanSidiropoulos2014}. This energy feedback based channel learning method is applicable for both TDD and FDD systems, and can  be implemented without additional baseband processing modules at the ER. However, the existing studies in \cite{XuZhang_OneBit,NoamGoldsmith2013,GopalakrishnanSidiropoulos2014} considered only one feedback bit per training interval; while it remains unknown how to efficiently implement this method for the general setup with multiple feedback bits available per interval. This thus motivates our study in this work.

In this paper, we propose a general channel learning design framework for MIMO WET based on the ER's energy feedback over a finite-rate reverse link from the ER to the ET (thus including the one-bit feedback per interval designed in \cite{XuZhang_OneBit} as a special case). Note that our results can also be similarly applied to other applications for channel learning in wireless communications with received signal power/energy based feedback (such as cognitive radio systems in \cite{NoamGoldsmith2013,GopalakrishnanSidiropoulos2014}). We consider a quasi-static channel model and a block-based energy transmission with each block spanning over a constant MIMO channel for WET.  Each block is further divided into two phases for channel learning and energy transmission, respectively. In the channel learning phase, the ET learns the MIMO channel progressively by adjusting its training beamforming weights according to the received energy feedback from the ER. It is assumed that at each feedback interval the ER sends $B \ge 1$ bits to the ET via the reverse link. 
In general, the feedback bits at each interval are generated by the ER via efficiently encoding the measured energy levels at the present and past intervals subject to the given feedback rate constraint. Based on the estimated MIMO channel from the channel learning phase, in the subsequent energy transmission phase, the ET implements the optimal transmit energy beamforming to maximize the energy transferred to the ER. In this paper, we focus our study on the channel learning phase by investigating the energy feedback design at the ER and the training beamforming design and channel estimation method at the ET. Our main results are summarized as follows.
\begin{itemize}
  \item First, we present a general design framework for the energy measurement feedback based channel learning with limited feedback rate. Similar to \cite{XuZhang_OneBit} for the special case of one-bit feedback (i.e., $B=1$), our design is based on the technique of analytic center cutting plane method (ACCPM) in convex optimization \cite{ACCPM}. The basic idea of ACCPM is as follows: Based on the $B$ feedback bits from the ER at each interval, the ET constructs one or more linear inequalities on the MIMO channel matrix to be estimated, where each inequality is used as a cutting plane for helping localize the MIMO channel in a multi-dimensional complex matrix space. Based on these cutting planes, the ET also designs the transmit beamforming for subsequent training intervals and obtains refined estimates of the MIMO channel.
  \item Under this general framework, we then propose two specific feedback designs, namely, energy quantization and energy comparison, where the $B$ feedback bits per interval are generated at the ER by quantizing the measured energy level at the current interval and comparing it with those in the $B$ previous intervals, respectively. Note that when $B=1$, the energy comparison based channel learning scheme degenerates to that with one-bit feedback per interval as in our previous work \cite{XuZhang_OneBit}.
  \item Finally, we provide extensive numerical results to compare the performance of our proposed feedback and channel learning schemes in terms of both mean-squared error of the estimated MIMO channel norm as well as the resulting energy beamforming gain. It is shown that energy quantization generally preforms better than energy comparison when $B$ is large, while energy comparison is more effective vice versa.
\end{itemize}

The remainder of this paper is organized as follows. Section \ref{sec:SysMod} introduces the system model and the two-phase transmission protocol. Section \ref{sec:Framework} presents the general design framework for the energy feedback based channel learning using ACCPM. Sections \ref{sec:quantization:feedback} and \ref{sec:comparison:feedback} present the energy quantization and energy comparison based feedback and channel learning schemes, respectively. Section \ref{sec:low:complexity} presents a complexity reduction approach for the proposed schemes by pruning irrelevant cutting planes in ACCPM. Section \ref{sec:numerical:results} provides numerical results to evaluate the performance of proposed schemes. Finally, Section \ref{sec:conclusion} concludes the paper.

{\it Notation:} Boldface letters refer to vectors (lower  case) or matrices (upper case). For a square matrix $\mv{S}$, $\det(\mv{S})$ and ${\mathrm{tr}}(\mv{S})$ denote its determinant and trace, respectively, while $\mv{S}\succeq \mv{0}$ means that $\mv{S}$ is positive semi-definite. For an arbitrary-size matrix $\mv{M}$, $\|\mv{M}\|_{\rm F}$, ${\mathrm{rank}}(\mv{M})$, $\mv{M}^H$, and $\mv{M}^T$ denote the Frobenius norm, rank, conjugate transpose and transpose of $\mv{M}$, respectively. $\mv{I}$ and $\mv{0}$ denote an identity matrix and an all-zero matrix, respectively, with appropriate dimensions. $\mathbb{C}^{x\times y}$ and $\mathbb{R}^{x\times y}$ denote the space of $x\times y$ complex and real matrices, respectively. ${\mathbb{E}}(\cdot)$ denotes the statistical expectation. $\|\mv{x}\|$ denotes the Euclidean norm of a complex vector $\mv{x}$, and $|z|$ denotes the magnitude of a complex number $z$. $\lfloor x \rfloor$  denotes the largest integer not greater than the real number $x$, and $\lceil x \rceil$ denotes the smallest integer not less than $x$, respectively. Symbol $j$ denotes the complex number $\sqrt{-1}$.

\section{System Model}\label{sec:SysMod}
We consider a point-to-point MIMO system for WET as shown in Fig. \ref{fig:System}, in which one ET with $M_T > 1$ transmit antennas sends wireless energy via RF signal to one ER with $M_R\ge 1$ receive antennas.{\footnote{Our framework is also applicable for distributed energy beamforming systems, where multiple distributed single-antenna ETs cooperatively send energy to one or more ERs provided that a central control unit (e.g., one of the ETs) is available to coordinate the transmission from multiple ETs by adopting the techniques proposed in this paper.}} We assume a quasi-static flat-fading channel model, in which the wireless channel remains constant over each block of our interest and may change from one block to another. Let the block length be denoted by $T > 0$, which is assumed to be sufficiently long for typical WET applications with low mobility receivers.

We consider transmit beamforming at the ET for WET. Without loss of generality, we assume that there are a total of $d$ energy beams employed at the ET, where $1\le  d\le M_T$ is a design parameter that will be specified later. We denote the transmitted energy signal at the ET as $\mv x = \sum_{i=1}^d \mv w_i s_i$, where $\mv w_i \in \mathbb{C}^{M_T \times 1}$ denotes the $i$th transmit beamforming vector and $s_i$ represents the corresponding energy-bearing signal. Since the energy signal $s_i$'s do not carry any information, they are assumed to be independent pseudorandom sequences with zero mean and unit variance, i.e., $\mathbb{E}(|s_i|^2) = 1, \forall i\in\{1,\ldots,d\}$, in order to spread the signal power over the frequency to
satisfy certain power spectral mask constraint imposed by the regulations on microwave radiation. Accordingly, the transmit covariance matrix at the ET is given by $\mv S = \mathbb{E}\left(\mv x\mv x^H\right) = \sum_{i=1}^d \mv w_i \mv w_i^H \succeq \mv 0$. Note that for any given transmit covariance matrix $\mv S$, the corresponding energy beamforming vectors, i.e., $\mv w_1, \ldots, \mv w_{d}$, can be obtained via performing the eigenvalue decomposition (EVD) on $\mv S$, where $d = \mathrm{rank}(\mv S)$. In addition, we assume that the maximum transmit sum-power of the ET is given by $P > 0$, and thus we have $\mathbb{E}(\|\mv x\|^2) = \mathrm{tr}\left(\mv S\right) \le P$.

With energy beamforming at the ET, the ER can harvest the energy carried by the $d$ energy beams from all the $M_R$ receive antennas. Let $\mv{H} \in \mathbb{C}^{M_R\times M_T}$ denote the MIMO channel matrix from the ET to the ER, and define $\mv{G}\triangleq {\mv{H}}^H\mv{H} \in \mathbb{C}^{M_T\times M_T}$. Due to the law of energy conservation, the harvested RF-band power from all receive antennas at the ER is proportional to o that of the equivalent baseband signal \cite{ZhangHo2013,ZhouZhangHo2013}. As a result, the harvested energy at the ER in one particular block of interest is expressed as
\begin{align}
Q = \eta T \mathbb{E}\left(\|\mv H \mv x\|^2\right) = \eta T \mathrm{tr}\left(\mv G \mv S\right),
\end{align}
where $0<\eta\le 1$ denotes the energy harvesting efficiency at the ER. Since $\eta$ is a constant, we normalize it as $\eta = 1$ in the sequel of this paper unless otherwise stated.

\subsection{Energy Transmission With and Without CSI}

In this subsection, we present the energy signal design with or without CSI at the ET. First, consider that the ET perfectly knows the CSI. In this case, the ET designs the energy beamforming to maximize its transferred energy to the ER subject to the given transmit sum-power constraint. Mathematically, the transferred energy maximization problem is formulated as
\begin{align}
\max\limits_{\mv S}~& T \mathrm{tr}\left(\mv G \mv S\right)\nonumber\\
\mathrm{s.t.}~& \mv S\succeq \mv 0, ~\mathrm{tr}\left(\mv S\right) \le P. \label{eqn:problem2}
\end{align}
Let the $M_T$ eigenvalues of $\mv G$ be denoted by $\lambda_1 \ge \ldots\ge \lambda_{M_T}\ge 0$, and the corresponding eigenvectors by $\mv v_{1},\ldots,\mv v_{M_T}$, respectively. Then it is shown in \cite{ZhangHo2013} that the optimal solution to problem (\ref{eqn:problem2}) is given by $\mv S^* = P \mv v_{1}\mv v_{1}^H$, which achieves the maximum transferred energy to the ER given by
\begin{align}\label{eqn:Qoeb}
Q^*  = T P \lambda_1.
\end{align}
Here, it follows that $\mathrm{tr}(\mv S^*) = 1$; as a result, only one single transmit energy beam is needed to achieve the optimality. Such a single-beam solution is termed as {\it optimal energy beamforming (OEB)}.

On the other hand, consider the case when the ET does not have any knowledge about the CSI. In this case, the OEB is not implementable. Alternatively, the ET can employ an isotropic transmission to radiate its transmit power uniformly over all directions, with the corresponding transmit covariance matrix given by $\mv S_{\rm iso} = P/{M_T}\mv I$. Correspondingly, the harvested energy at the ER is expressed as
\begin{align}\label{eqn:Qiso}
Q_{\rm iso} = T \mathrm{tr}\left(\mv G \mv S_{\rm iso}\right) = \frac{T P\mathrm{tr}(\mv G)}{M_T} = \frac{TP{\sum_{i=1}^{M_T}\lambda_i}}{M_T}.
\end{align}

By comparing $Q^*$ in (\ref{eqn:Qoeb}) and $Q_{\rm iso}$ in (\ref{eqn:Qiso}), it is observed that the OEB with perfect CSI at the ET offers an energy beamforming gain over the isotropic transmission without CSI, which can be expressed as
\begin{align}\label{eqn:EBGain}
\chi^* = \frac{Q^*}{Q_{\rm iso}} = \frac{M_T\lambda_1}{\mathrm{tr}(\mv G)}= \frac{M_T\lambda_1}{\sum_{i=1}^{M_T}\lambda_i}.
\end{align}
Due to the fact that $\lambda_1 \ge \ldots\ge \lambda_{M_T}\ge 0$, it is evident that $1\le \chi^* \le M_T$, where $\chi^*  = 1$ holds when $\lambda_1 = \cdots = \lambda_{M_T} > 0$ (with $\mathrm{rank}(\mv G) = M_T$), while $\chi^*= M_T$ is attained when $\lambda_1 > 0 $ and $\lambda_2 =\cdots = \lambda_{M_T} = 0$ (i.e., $\mathrm{rank}(\mv G) = 1$). For practical WET systems with $M_T \ge M_R$, we normally have $\lambda_1 \gg \lambda_2 \ge \cdots\ge  \lambda_{M_R}> 0$ (and $\lambda_{M_R+1} = \cdots =  \lambda_{M_T} = 0$ if $M_T > M_R$), since the wireless channel is often line-of-sight (LOS) dominant given the limited transmission range of most interest (say, several meters). In this case, an energy beamforming gain $\chi^* \approx M_T$ can be achieved with the availability of CSI at the ET.

\subsection{Two-Phase Transmission Protocol}

\begin{figure}
\centering
 \epsfxsize=1\linewidth
    \includegraphics[width=8cm]{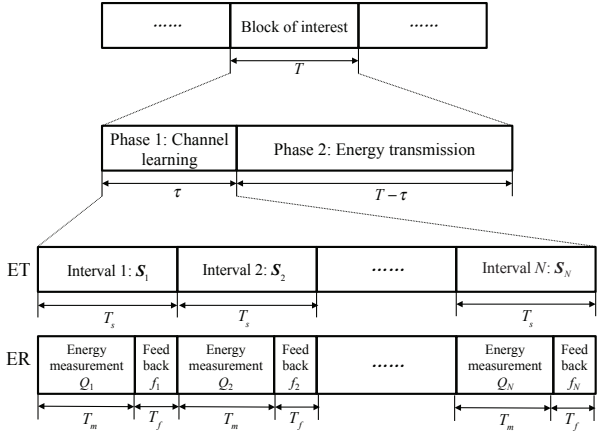}
\caption{The two-phase transmission protocol for MIMO WET.} \label{fig:protocol}
\end{figure}

To achieve the promising energy beamforming gain in practice, we consider a two-phase transmission protocol for the ET to learn the MIMO channel and based on the estimated channel implement the OEB. As shown in Fig. \ref{fig:protocol}, each transmission block is divided into two phases, for the purposes of channel learning and energy transmission, respectively. We explain these two phases in detail as follows.

First, the channel learning phase corresponds to the first $\tau$ amount of time in each block, $0 < \tau < T$, which is further divided into $N \geq 1$ training intervals. Let $T_s$ denote the length of each training interval. Then we have $\tau = NT_s$. In each of the $N$ intervals, the ET transmits different sets of energy beams (each set corresponding to a different transmit covariance matrix) to the ER. Let the transmit covariance matrix at the ET in the interval $n\in\{1,\ldots,N\}$ be denoted by $\mv S_{n}$, where $\mv S_n \succeq \mv 0$ and $\mathrm{tr}(\mv S_n) \le P$. Then the harvested energy by the ER in the $n$-th interval is given by
\begin{align}\label{eqn1}
Q_n^{\rm CL} = T_s\mathrm{tr}(\mv G\mv S_n), n\in\{1,\ldots,N\}.
\end{align}
Specifically, the ET implements energy measurement via the energy meter shown in Fig. \ref{fig:System} in the first $T_m$ amount of time in each interval,\footnote{Alternatively, the ER can use multiple energy meters each for measuring the harvested energy at one receive antenna. In this case, the MIMO channel learning becomes estimating an $M_R$ number of multiple-input single-output (MISO) channels each for one receive antenna. This can increase the channel learning convergence rate at the cost of higher hardware and feedback costs at the ER.} and then sends back the measured energy information to the ET at the remaining $T_f$ amount of time, where $T_m + T_f = T_s$. The harvested energy amount measured by the ER in the $n$-th interval is denoted by
\begin{align}\label{eqn1:measured}
Q_n = T_m\mathrm{tr}(\mv G\mv S_n), n\in\{1,\ldots,N\}.
\end{align}
It is assumed that the ER can send $B$ bits to the ET at each interval via the reverse link, where $B =  \lfloor RT_f \rfloor \ge 1$ is  an integer with $R$ denoting the feedback rate from the ER to the ET. The value of $B$ is practically finite since the ER is energy-constrained for sending training signals and the available bandwidth for feedback may be shared among multiple ERs (to their respective ETs). We denote the $B$-bits feedback at each interval $n$ as $f_n$. In general, $f_n$ specifies the energy feedback scheme in use at the ER which is a mapping from its present and past energy observations (i.e., $Q_1,\ldots, Q_n$) to the $B$ feedback bits at interval $n, n\in\{1,2,\ldots,N\}$. Furthermore, let $E_{m} > 0$ and $E_{f} > 0$ denote the energy consumed by the ER for the energy measurement and the information feedback in each interval, respectively. Note that the values of $T_m$ and $T_f$ should be chosen in practice by considering the sensitivity of the energy meter at each ER and the feedback link capacity from each ER to the ET, respectively, as well as the ER's energy consumption for energy measurement and feedback.

For the purpose of exposition, we make the following assumptions on the ER's energy measurement and feedback.
\begin{itemize}
\item The energy measurement of each $Q_n$ at the ER is perfectly accurate, while the effect of energy measurement error at the ER on the channel learning performance will be shown by simulation in Section \ref{sec:EnergyMeasurementError}.
\item The feedback link is ideal and each feedback $f_n$ is received by the ET at the end of interval $n$ without any error or delay;
\end{itemize}
Based on the feedback $\{f_n\}_{n=1}^N$, the ET obtains a final estimate of the MIMO channel at the end of the channel learning phase. The details of energy feedback design at the ER and training signal design and channel estimation method at the ET will be given later.

Next, in the energy transmission phase with the remaining time $T-\tau$, the ET implements the OEB for WET based on the estimated MIMO channel from the previous channel learning phase. Let the dominant eigenvector of the estimated MIMO channel be denoted by $\tilde{\mv v}_1$. Then the ET sets the transmit covariance matrix in the energy transmission phase as
\begin{align}\label{eqn:OEB}
\mv S_{\rm ET} = P \mv {\tilde v}_{1} \mv {\tilde v}_{1}^H.
\end{align}
Accordingly, the transferred energy to the ER is expressed as
\begin{align}
Q_{\rm ET} = P (T-\tau) \mv {\tilde v}_{1}^H \mv G \mv {\tilde v}_{1}.
\end{align}

By combing the channel learning and energy transmission phases, the net energy harvested by the ER over one whole block is expressed as its totally harvested energy minus the total energy consumed for energy measurement and feedback, i.e.,
\begin{align}\label{eqn:tildeQ}
Q_{\rm net} = \sum_{n=1}^N T_s\mathrm{tr}(\mv G\mv S_n)+ P(T-\tau) \mv {\tilde v}_{1}^H \mv G \mv {\tilde v}_{1} - N \left(E_m + E_f \right).
\end{align}
Accordingly, we define the practical energy beamforming gain achieved by the two-phase protocol as
\begin{align}
\chi = \frac{Q_{\rm net}}{Q_{\rm iso}}.
\end{align}
For typical WET applications with $T$ being sufficiently large, it follows that $\tau \ll T$ and $\sum_{n=1}^N T_s\mathrm{tr}(\mv G\mv S_n)
\ll P (T-\tau) \mv {\tilde v}_{1}^H \mv G \mv {\tilde v}_{1}$. Furthermore, it can be assumed that the energy consumption $E_m$ and $E_f$ are negligible as compared to the totally harvested energy, i.e., $N \left(E_m + E_f \right) \ll P (T-\tau) \mv {\tilde v}_{1}^H \mv G \mv {\tilde v}_{1}$.\footnote{To maximize the net energy harvested at the ER, there in general exists a design tradeoff in choosing $\tau$ versus $T-\tau$ between channel learning and energy transmission. Due to page limitation, we omit such discussion here and the interested readers can refer to \cite{ZengZhang2014A,ZengZhang2014B,YangHoZhangGuan2014,YangHoGuan2014,XuZhang_OneBit} for details.} Therefore, we approximate the practical energy beamforming gain as
\begin{align}\label{eqn:chiEB}
\chi \approx \frac{P (T-\tau) \mv {\tilde v}_{1}^H \mv G \mv {\tilde v}_{1} }{PT\mathrm{tr}(\mv G)/M_T} \approx  \tilde\chi\triangleq \frac{M_T \mv {\tilde v}_{1}^H \mv G \mv {\tilde v}_{1}}{\mathrm{tr}(\mv G)}.
\end{align}
For convenience, in the sequel of this paper we use the approximate energy beamforming gain $\tilde\chi$ as the performance metric for proposed feedback and channel learning schemes. Note that when the channel estimation is sufficiently accurate, i.e., $\tilde{\mv v}_1 \approx \mv v_1$, the energy beamforming gain $\tilde\chi$ approaches that by the OEB with perfect CSI at the ET $\chi^*$ in (\ref{eqn:EBGain}), i.e., $\tilde\chi \approx \chi^*$.

\section{A General Channel Learning Design Framework Based On ACCPM}\label{sec:Framework}

In this section, we present a general framework for designing the energy feedback at the ER and the corresponding training signal and channel estimation at the ET based on ACCPM. ACCPM is well known as an efficient localization and cutting plane method for solving convex or quasi-convex optimization problems \cite{ACCPM}, with the objective of finding one feasible point in a convex target set, where the target set can be the set of optimal solutions to the problem of interest. The implementation of ACCPM is based on the use of cutting planes, each of which is a hyperplane that separates some of the current points from the points in the target set. Based on the returned cutting planes in each iteration, the target set can be localized via a sequence of convex working sets with reduced size.

To implement ACCPM in our context, we first define the target set for channel learning as follows. Note that it follows from (\ref{eqn:OEB}) that the implementation of OEB at the ET in the energy transmission phase only depends on $\tilde{\mv v}_1$, which is the estimate of the dominant eigenvector of the matrix $\mv G=\mv H^H\mv H$. For convenience, in the sequel we denote $\mv G$ as the MIMO channel to be estimated, instead of the actual MIMO channel $\mv H$. To obtain $\tilde{\mv v}_1$, it suffices for the ET to estimate any positively scaled matrix of $\mv G$ in the channel learning phase. Without loss of generality, we focus on learning the normalized MIMO channel given by $\bar{\mv G} \triangleq {\mv G}/{\mathrm{tr}(\mv G)}$ with $\mathrm{tr}(\bar{\mv G}) = 1$. Accordingly, the target set of our interest is defined as $\varphi \triangleq \left\{\bar{\mv G}\right\}$.

Next, we explain the principle for designing the feedback $\{f_n\}$, and show that they in fact play the role of specifying the cutting planes in ACCPM for localizing $\varphi$. To facilitate the design of $\{f_n\}$, we first set the transmit covariance matrix at the ET in the first interval $n=1$ to be $\mv S_{1} = P/{M_T}\mv I$ (i.e., isotropic transmission), based on which the corresponding harvested energy amount measured by the ER is $Q_1 = T_m\mathrm{tr}(\mv{GS}_1) = T_mP/M_T\mathrm{tr}(\mv{G})$. By using $Q_1$ as a reference energy level, we then normalize the harvested energy measurement in each interval $n \in \{1,\ldots,N\}$ as
\begin{align}\label{eqn:barQ}
\bar Q_n = \frac{Q_n}{Q_1} = \frac{M_T\mathrm{tr}(\mv{GS}_n) }{P\mathrm{tr}(\mv{G})}= \frac{M_T}{P}\mathrm{tr}\left(\bar{\mv G} \mv S_n\right).
\end{align}
Note that for interval $n=1$, we have $\bar Q_1=\mathrm{tr}(\bar{\mv G}) =1$; as a result, the feedback over this interval $f_1$ will not provide any information on $\bar{\mv G}$ and thus can be  ignored. This assumption is made specifically for our proposed channel learning algorithms, while in general it may not be optimal for other designs. Since $\bar Q_n, n\in\{2,\ldots,N\}$  is a real number but $f_n$ is of $B$ bits only, each feedback $f_n$ cannot completely convey $\bar Q_n$ to the ET at interval $n$; therefore, $f_n$ should be designed by properly encoding the present and past energy measurements, i.e., $\bar{Q}_2,\ldots, \bar Q_n$, at each interval $n\in\{2,\ldots,N\}$. The optimal encoding scheme subject to $B$ feedback bits per interval is a difficult problem to solve; thus we will later propose two suboptimal designs for efficient practical implementation in Sections \ref{sec:quantization:feedback} and \ref{sec:comparison:feedback}, respectively. For each proposed feedback scheme, a set of new linear inequalities (or cutting planes in ACCPM) on $\bar{\mv G}$ are obtained from $f_n$ at each interval $n \in \{2,\ldots,N\}$, which are generally expressed as
\begin{align}\label{eqn:general:cut}
\mathrm{tr}\left(\mv \Sigma_{n,c} \bar{\mv G}\right) - \gamma_{n,c}  \le 0, c\in\{1,\ldots,C_n\},
\end{align}
where $C_n \ge 1$ denotes the number of inequalities/cutting planes obtained at interval $n$, and $\{\mv \Sigma_{n,c}\}$ and $\{\gamma_{n,c}\}$ are parameters depending on the design of transmit covariance matrix $\mv S_n$'s and feedback $f_n$'s in each scheme (to be specified later). Each of the $C_n$ cutting planes in (\ref{eqn:general:cut}) ensures that the MIMO channel $\bar {\mv G}$ should lie in the half space of
\begin{align}
\mathcal H_{n,c} = &\left\{\hat{\mv G}\big|\mathrm{tr}\left(\mv \Sigma_{n,c} \hat{\mv G}\right) - \gamma_{n,c}  \le 0 \right\}, \nonumber\\
&~~c\in\{1,\ldots,C_n\}, n \in \{2,\ldots,N\}.
\end{align}

Now, based on the returned cutting planes in (\ref{eqn:general:cut}), we are ready to obtain a sequence of working sets to localize the target set $\varphi$. Since $\bar{\mv G}\succeq \mv 0$ and $\mathrm{tr}(\bar{\mv G}) = 1$ are {\it a priori} known by the ET, we have the initial working set as $\mathcal{P}_0 = \left\{\hat{\mv G}\big|\hat{\mv G} \succeq\mv{0}, \mathrm{tr}(\hat{\mv G}) = 1 \right\}$, with $\mathcal{P}_0 \supseteq \varphi$. Furthermore, for the first interval $n=1$, we have the working set $\mathcal{P}_1=\mathcal{P}_0$, since $\bar Q_1=\mathrm{tr}(\bar{\mv G}) = 1$ is a constant and $f_1$ does not contain any information on $\bar{\mv G}$. Then, for the subsequent intervals $n\in\{2,\ldots,N\}$, the working set can be updated as $\mathcal P_n = \mathcal P_{n-1} \cap \mathcal H_{n,1} \cdots \cap \mathcal H_{n,C_n}$, or more explicitly,
\begin{align}\label{eqn:working:set}
\mathcal P_n = \bigg\{\hat{\mv G}\big|&\hat{\mv G} \succeq \mv 0, \mathrm{tr}\left(\hat {\mv G}\right)=1, \mathrm{tr}\left(\mv \Sigma_{m,c} \hat{\mv G}\right) - \gamma_{m,c}  \le 0, \nonumber\\
&\forall c\in\{1,\ldots,C_m\}, \forall m\in\{2,\ldots,n\} \bigg\}.
\end{align}
It is evident that $\mathcal{P}_0 =\mathcal{P}_1 \supseteq \cdots \supseteq \mathcal{P}_N \supseteq \varphi$. Therefore, the sequence of obtained working set $\mathcal{P}_n$'s shrink in size and eventually converge to $\varphi$ as $N\to\infty$. Here, the convergence of our proposed ACCPM based channel learning can be proved, similarly as that given in \cite{XuZhang_OneBit} for the special case of $B=1$. For brevity, we omit the proof in this paper.

\begin{figure}
\centering
 \epsfxsize=1\linewidth
    \includegraphics[width=8cm]{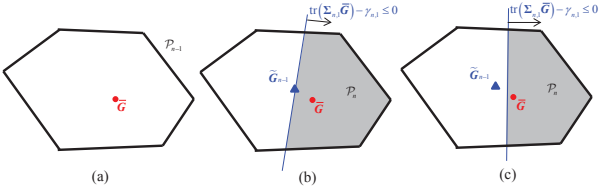}
\caption{Illustration of using neutral and deep cutting planes to update the working set at interval $n\in\{2,\ldots,N\}$ with ACCPM: (a) the working set $\mathcal P_{n-1}$ containing the MIMO channel $\bar{\mv G}$ to be learned; (b) a neutral cutting plane that passes through $\tilde{\mv G}_{n-1}$, which is the analytic center of $\mathcal P_{n-1}$, to cut nearly a half space of $\mathcal P_{n-1}$, with $\mathrm{tr}\left(\mv \Sigma_{n,1} \tilde{\mv G}_{n-1}\right) - \gamma_{n,1}  = 0$; (c) a deep cutting plane that locates between $\bar{\mv G}$ and $\tilde{\mv G}_{n-1}$ to cut a more substantial sub-space of $\mathcal P_{n-1}$, with $\mathrm{tr}\left(\mv \Sigma_{n,1} \tilde{\mv G}_{n-1}\right) - \gamma_{n,1}  > 0$.} \label{fig:framework:ACCPM}
\end{figure}

Finally, to complete the ACCPM based channel learning, it remains to design the cutting planes in (\ref{eqn:general:cut}) in terms of $\{\mv \Sigma_{n,c}\}$ and $\{\gamma_{n,c}\}$, $c\in\{1,\ldots,C_n\}$, for each interval $n\in\{2,\ldots,N\}$ via adjusting the transmit covariance matrix $\mv S_n$. Consider a feedback interval $n$ with the given working set $\mathcal P_{n-1}$ as shown in Fig. \ref{fig:framework:ACCPM}(a). Then, based on the principle of ACCPM \cite{ACCPM}, a desired cutting plane should be {\it neutral} or {\it deep}. A neutral cutting plane should pass through the so-called analytic center of $\mathcal P_{n-1}$, denoted by $\tilde{\mv{G}}_{n-1}$, so that nearly a half portion of $\mathcal P_{n-1}$ can be ``cut'' or eliminated, as shown in Fig. \ref{fig:framework:ACCPM}(b); while a deep cutting plane should locate between the analytic center $\tilde{\mv G}_{n-1}$ and the MIMO channel $\bar{\mv G}$ to be learned, such that a more substantial sub-space of $\mathcal P_{n-1}$ can be cut as compared to a neural cutting plane, as shown in Fig. \ref{fig:framework:ACCPM}(c). Here, for each convex working set $\mathcal P_n$, its analytic center is defined as
\begin{align}
\tilde{\mv{G}}_n =& \mathrm{arg}\mathop\mathrm{min}\limits_{\hat{\mv G}\succeq \mv 0, \mathrm{tr}(\hat{\mv G}) = 1} -\log\det\left(\hat{\mv G}\right)- \nonumber\\
&\sum_{m=2}^n\sum_{c=1}^{C_m}\log\left(-\mathrm{tr}\left(\mv \Sigma_{m,c} \hat{\mv G}\right) + \gamma_{m,c}\right), n\in \{1,\ldots,N\}.\label{eqn:accpm:general}
\end{align}
Note that the minimization problem in (\ref{eqn:accpm:general}) can be shown to be convex \cite{BV:Convex}, and thus is solvable by standard convex optimization techniques such as CVX \cite{cvx}. Also note that we have $\tilde{\mv{G}}_1 = \mv I/M_T$ from (\ref{eqn:accpm:general}). As a result, in order for the $c$th cutting plane to be neutral or deep (see Fig. \ref{fig:framework:ACCPM}), we design the transmit covariance matrix $\mv S_n$ to ensure
\begin{align}\label{eqn:cut:general}
\mathrm{tr}\left(\mv \Sigma_{n,c} \tilde{\mv G}_{n-1}\right) - \gamma_{n,c}  \ge 0, c\in\{1,\ldots,C_n\}.
\end{align}
Note that when $C_n > 1$, it may be difficult to guarantee all the $C_n$ cutting planes to be neutral or deep. In such cases, we can select any one of them to be the neutral/deep cutting plane to satisfy (\ref{eqn:cut:general}). The detailed design of such $\mv S_n$'s will be explained later for each specific feedback scheme to be introduced.

In summary, a generic algorithm for channel learning based on ACCPM is described as Algorithm 1 in Table I. Note that at the last interval $N$, the ET adopts the analytic center $\tilde{\mv G}_N$ as the final estimate of the MIMO channel in the channel learning phase, given by $\tilde{\mv G} = \tilde{\mv G}_N$. Under the above general design framework with ACCPM, in the next two sections, we present two specific feedback schemes for channel learning, respectively.

\begin{table}[!t]\scriptsize
\caption{A Generic Algorithm for Channel Learning based on ACCPM }
\label{table:framework} \centering
\begin{tabular}{|p{8.5cm}|}
\hline
\textbf{Algorithm 1}\\
\hline\vspace{0.01cm}
1) {\bf Initialization:} Set $n=0$ and $\mv S_1 = P/{M_T}\mv I$. \\
2) {\bf Repeat:}
  \begin{itemize} \setlength{\itemsep}{0pt}
    \item[a)] $n \gets n+1$;
    \item[b)] The ET transmits with $\mv{S}_n$;
    \item[c)] The ER determines $f_n$ (if $n\geq 2$) based on the normalized energy measurements, $\bar Q_1,\ldots,\bar Q_n$ in (\ref{eqn:barQ}), and then sends $f_n$ back to the ET;
    \item[d)] The ET extracts a set of $C_n$ cutting planes given in (\ref{eqn:general:cut}) based on $f_n$ (if $n\geq 2$), and computes the analytic center $\tilde{\mv G}_n$ as in (\ref{eqn:accpm:general});
    \item[e)] The ET designs $\mv S_{n+1}$ to ensure $\mathrm{tr}(\mv \Sigma_{n+1,c} \tilde{\mv G}_{n}) - \gamma_{n+1,c}  \geq 0$ for some $c\in\{1,\ldots,C_n\}$.
  \end{itemize}
  3) {\bf Until}  $n\ge N$.\\
  4) The ET obtains the channel estimate as $\tilde{\mv G}=\tilde{\mv G}_N$.\\
 \hline
\end{tabular}
\end{table}

\section{Energy Quantization Based Feedback and Channel Learning}\label{sec:quantization:feedback}

In this section, we propose an energy quantization based feedback scheme, where at each interval $n$ the ER obtains the feedback $f_n$ by quantizing its normalized energy measurement $\bar Q_n$ at the current interval into $B$ bits. For this scheme, we denote the feedback at the ER as $\left\{f_n^{(1)}\right\}$, the transmit covariance matrices at the ET as $\left\{\mv S_n^{(1)}\right\}$, the updated sequence of analytic centers as $\left\{\tilde{\mv G}_n^{(1)}\right\}$, and the estimated MIMO channel as $\tilde{\mv G}^{(1)}$.

\subsection{Feedback Design at ER}

First, we present the feedback design for $\left\{f_n^{(1)}\right\}$ at the ER by quantizing the normalized energy measurements $\left\{\bar Q_n\right\}$. Among various quantization methods \cite{Quantization}, we adopt a simple uniform scalar quantizer that is of low complexity for energy-efficient implementation at the ER. Consider an interval $n \in \{2,\ldots,N\}$, and assume that the normalized energy measurement $\bar Q_n$ is within the unit interval $(0,1]$. {{Note that if an overflow (i.e., $\bar Q_n > 1$) occurs, then we set $\bar Q_n = 1$. To minimize the probability of overflow, we should properly set the transmit power at the ET as will be specified later (see (\ref{neutral:1})).}} Then, the uniform quantizer maps $\bar Q_n$ into a discrete variable that belongs to a finite set, given by $\left\{\frac{\Delta}{2}+i\Delta\big|i=0,\ldots,2^B-1\right\}$, where there are in total $2^B$ quantization levels resulted from the $B$ feedback bits, and $\Delta = 2^{-B}$ denotes the quantization step size. As a result, the quantized value is expressed as
\begin{align}\label{eqn:quantize}
\tilde Q_n = \frac{\left\lceil2^B\bar Q_{n}\right\rceil}{2^B} - \frac{1}{2^{B+1}},
\end{align}
where $\lceil x\rceil$ denotes the minimum integer that is no smaller than $x$. Accordingly, by encoding the quantized value $\tilde Q_n$ as a $B$-bits digital codeword, the feedback $f_n^{(1)}$ can be obtained, which is then sent back to the ET.

After receiving the feedback $f_n^{(1)}$, $n\in\{2,\ldots,N\}$, the ET has the knowledge of the quantized value $\tilde Q_n$, and accordingly knows that the normalized energy measurement $\bar Q_n$ should lie between two thresholds $\tilde Q_n - \frac{1}{2^{B+1}}$ and $\tilde Q_n + \frac{1}{2^{B+1}}$, i.e.,{\footnote{It is worth noting that if $\tilde Q_n = \frac{1}{2^{B+1}}$, then the first inequality in (\ref{eqn:inequations:1}) becomes $\bar Q_n \ge 0$, which is known {\it a priori} and thus is redundant. On the other hand, if $\tilde Q_n = 1-\frac{1}{2^{B+1}}$, then the second inequality in (\ref{eqn:inequations:1}) becomes $\bar Q_n \le 1$, which, however, may be unreliable due to the possibility of overflow. In the above two cases, the corresponding inequalities should be discarded.}}
\begin{align}\label{eqn:inequations:1}
\tilde Q_n - \frac{1}{2^{B+1}} \le \bar Q_n \le \tilde Q_n + \frac{1}{2^{B+1}}.
\end{align}
By using (\ref{eqn:inequations:1}) together with (\ref{eqn:barQ}), it follows that
\begin{align}
- \frac{M_T}{P}\mathrm{tr}\left(\bar{\mv G} \mv S_n^{(1)}\right) + \tilde Q_n - \frac{1}{2^{B+1}} &\le 0,\label{eqn:inequations:2:1}\\
\frac{M_T}{P}\mathrm{tr}\left(\bar{\mv G} \mv S_n^{(1)}\right) - \tilde Q_n - \frac{1}{2^{B+1}} &\le 0.\label{eqn:inequations:2}
\end{align}
The two inequalities in (\ref{eqn:inequations:2:1}) and (\ref{eqn:inequations:2}) correspond to $C_n=2$ cutting planes in the general expression (\ref{eqn:general:cut}) to localize the MIMO channel $\bar{\mv G}$ with $\mv \Sigma_{n,1} =- M_T\mv S_n^{(1)}/P$, $\gamma_{n,1} =- \tilde Q_n + \frac{1}{2^{B+1}}$, $\mv \Sigma_{n,2} = M_T\mv S_n^{(1)}/P$, and $\gamma_{n,2} = \tilde Q_n + \frac{1}{2^{B+1}}$.

\subsection{Training Signal Design and Channel Estimation at ET}\label{sec:comparison:feedback:2}

Based on the feedback $f_n^{(1)}$'s, the ET obtains the working set for the interval $n\in\{1,\ldots,N\}$ similarly as in (\ref{eqn:working:set}) by replacing the cutting planes in (\ref{eqn:general:cut}) as those in (\ref{eqn:inequations:2:1}) and (\ref{eqn:inequations:2}), i.e.,
 \begin{align}
\mathcal P_n^{(1)}& = \bigg\{\hat{\mv G}\big|\hat{\mv G} \succeq \mv 0, \mathrm{tr}\left(\hat {\mv G}\right)=1, \nonumber\\
 &- \frac{M_T}{P}\mathrm{tr}\left(\hat{\mv G} \mv S_m^{(1)}\right) + \tilde Q_m - \frac{1}{2^{B+1}} \le 0,  \nonumber\\& \frac{M_T}{P}\mathrm{tr}\left(\hat{\mv G} \mv S_m^{(1)}\right) - \tilde Q_m - \frac{1}{2^{B+1}}\le 0, \forall m\in\{2,\ldots,n\}\bigg\}.\label{eqn:working:set:1}
\end{align}
Accordingly, the analytic center of $\mathcal P_n^{(1)}$ is expressed as
\begin{align}\nonumber
\tilde{\mv{G}}_n^{(1)} = &\mathrm{arg}\mathop\mathrm{min}\limits_{\hat{\mv G}\succeq \mv 0, \mathrm{tr}(\hat{\mv G}) = 1} -\log\det\left(\hat{\mv G}\right) \nonumber\\& -\sum_{m=2}^n\log\left(\frac{M_T}{P}\mathrm{tr}\left(\hat{\mv G}\mv S_m^{(1)}\right) - \tilde Q_m + \frac{1}{2^{B+1}}\right)\nonumber \\&-\sum_{m=2}^n\log\left(\tilde Q_m+ \frac{1}{2^{B+1}} -\frac{M_T}{P}\mathrm{tr}\left(\hat{\mv G}\mv S_m^{(1)}\right)  \right).\label{eqn:accpm}
\end{align}

Finally, to complete the energy quantization based scheme, we design the transmit covariance matrix $\mv S_n^{(1)}$ at each interval $n \in \{2,\ldots,N\}$ to make one of the cutting planes in (\ref{eqn:inequations:2:1}) and (\ref{eqn:inequations:2}) to be neutral or deep with respect to $\tilde{\mv G}_{n-1}^{(1)}$. Towards this end, we design $\mv S_n^{(1)}$ to ensure that
\begin{align}
\frac{M_T}{P}\mathrm{tr}\left(\tilde{\mv{G}}_{n-1}^{(1)} \mv S_n^{(1)}\right) &= \frac{1}{2}, \label{neutral:1}
\end{align}
where $\frac{1}{2}$ corresponds to the threshold between the two consecutive quantization levels $\frac{1}{2} - \frac{1}{2^{B+1}}$ and $\frac{1}{2}+\frac{1}{2^{B+1}}$. From (\ref{neutral:1}), it follows that when $\tilde Q_n \ge \frac{1}{2} + \frac{1}{2^{B+1}}$, we have $-\frac{M_T}{P}\mathrm{tr}\left(\tilde{\mv G}_{n-1}^{(1)} \mv S_n^{(1)}\right) + \tilde Q_n - \frac{1}{2^{B+1}} \ge 0$, and hence the cutting plane in (\ref{eqn:inequations:2:1}) is neutral or deep; otherwise (when $\tilde Q_n \le \frac{1}{2} - \frac{1}{2^{B+1}}$), we have $\frac{M_T}{P}\mathrm{tr}\left(\tilde{\mv G}_{n-1}^{(1)} \mv S_n^{(1)}\right) - \tilde Q_n - \frac{1}{2^{B+1}} \ge 0$, and thus the cutting plane in (\ref{eqn:inequations:2}) is also neutral or deep. Thus, both cases satisfy the requirement given in (\ref{eqn:cut:general}) for ACCPM.  In particular, at each interval $n \in \{2,\ldots,N\}$, we set the transmit covariance matrix as
\begin{align}\label{eqn:S:LSM}
\mv S_n^{(1)} = p_n{\mv A_n^H \mv A_n},
\end{align}
where $\mv A_n$ is an $M_T\times M_T$ complex matrix, each element of which follows a circularly symmetric complex Gaussian (CSCG) distribution with zero mean and unit variance, and to satisfy (\ref{neutral:1}) $p_n>0$ is given by
\begin{align}
p_n = \frac{P}{2M_T\mathrm{tr}\left(\tilde{\mv{G}}_{n-1}^{(1)} \mv A_n^H \mv A_n\right)}.
\end{align}
Here, the randomness of $\mv A_n$ in $\mv S_n^{(1)}$ helps make the cutting planes over different feedback intervals as diverse as possible, so that they are more effective in cutting. Furthermore, several random trials may be required to make sure that $\mathrm{tr}\left(\mv S_n^{(1)}\right) \le P$.

In summary, the energy quantization based feedback and channel learning scheme is described as Algorithm 2 in Table \ref{table:2:1}.

\begin{table}[!t]\scriptsize
\caption{Energy Quantization Based Scheme}
\label{table:2:1} \centering
\begin{tabular}{|p{8.5cm}|}
\hline
\textbf{Algorithm 2}\\
\hline\vspace{0.01cm}
1) {\bf Initialization:} Set $n=0$ and $\mv S_1^{(1)} = P/{M_T}\mv I$. \\
2) {\bf Repeat:}
  \begin{itemize} \setlength{\itemsep}{0pt}
    \item[a)] $n \gets n+1$;
    \item[b)] The ET transmits with $\mv{S}_n^{(1)}$;
    \item[c)] The ER determines $f_n^{(1)}$ (if $n\ge 2$) by encoding the quantized value $\tilde Q_n$ in (\ref{eqn:quantize}) as a $B$-bits digital codeword, and then sends $f_n^{(1)}$ back to the ET;
    \item[d)] The ET updates the working set as $\mathcal P_n^{(1)}$ in (\ref{eqn:working:set:1}), and computes the analytic center of $\mathcal P_n^{(1)}$ as $\tilde{\mv G}_{n}^{(1)}$ given in (\ref{eqn:accpm});
    \item[e)] The ET generates $\mv S_{n+1}^{(1)}$ given in (\ref{eqn:S:LSM}).
  \end{itemize}
  3) {\bf Until}  $n\ge N$.\\
  4) The ET obtains the channel estimate as $\tilde{\mv G}^{(1)}=\tilde{\mv G}^{(1)}_N$.\\
 \hline
\end{tabular}
\end{table}

\subsection{Special Case: $B \to \infty$}\label{sec:perfect:feedback}

It is interesting to discuss the energy quantization based scheme in the extreme case of unlimited number of feedback bits per interval, i.e., $B\to\infty$. In this ideal case, we have $\frac{1}{2^{B+1}}\to 0$, and the two inequalities in (\ref{eqn:inequations:2:1}) and (\ref{eqn:inequations:2}) can be combined into one single linear equality as
\begin{align}
\frac{M_T}{P}\mathrm{tr}\left(\bar{\mv G} \mv S_n^{(1)}\right) - \tilde Q_n = 0,n\in\{2,\ldots,N\}.\label{eqn:inequations:3}
\end{align}
As a consequence, the computation of the analytic center in (\ref{eqn:accpm}) at the interval $n\in\{1,\ldots,n\}$ can be re-expressed as
\begin{align}\nonumber
\tilde{\mv{G}}_n^{(1)} = &\mathrm{arg}~\mathop\mathrm{min}\limits_{\hat{\mv G}} -\log\det\left(\hat{\mv G}\right)\\
&\mathrm{s.t.}~\frac{M_T}{P}\mathrm{tr}\left(\hat{\mv G} \mv S_m^{(1)}\right) - \tilde Q_m = 0,\forall m\in\{2,\ldots,n\}\nonumber\\
&~~~~~\hat{\mv G}\succeq \mv 0, \mathrm{tr}\left(\hat{\mv G}\right) = 1.\label{eqn:accpm:B}
\end{align}
To draw insights from problem (\ref{eqn:accpm:B}), we define a vector operation $\mathrm{cvec}(\cdot)$ as follows. For any complex Hermitian matrix $\mv{X} \in \mathbb{C}^{z \times z}$ that contains $z^2$ independent real elements, the vector operation $\mathrm{cvec}(\cdot)$ maps $\mv{X}$ to a real vector $\mathrm{cvec}(\mv{X}) \in \mathbb{R}^{z^2 \times 1}, z\ge1$, where all elements of $\mathrm{cvec}(\mv{X})$ are independent from each other, and $\mathrm{tr}({\mv{X}}{\mv{Y}}) = (\mathrm{cvec}(\mv{X}))^T\mathrm{cvec}(\mv{Y})$ for any given complex Hermitian matrix $\mv{Y}$.\footnote{The mapping between the complex Hermitian matrix $\mv{X} \in \mathbb{C}^{z \times z}$ and the real vector $\mathrm{cvec}(\mv{X})  \in \mathbb{R}^{z^2 \times 1}$, $z\ge 1$, can be realized as follows. The first $z$ elements of $\mathrm{cvec}(\mv{X}) $ consist of the diagonal elements of $\mv{X}$ (that are real), i.e., $[\mv{X}]_{aa}$'s, $\forall a\in\{1,\ldots,z\}$, where $[\mv{X}]_{ab}$ denotes the element in the $a$th row and $b$th column of $\mv{X}$; the next $\frac{z^2-z}{2}$ elements of $\mathrm{cvec}(\mv{X}) $ are composed of the (scaled) real part of the upper (or lower) off-diagonal elements of $\mv{X}$, i.e., $\frac{[\mv{X}]_{ab} + [\mv{X}]_{ba}}{\sqrt{2}}$'s, $\forall a,b\in\{1,\ldots,z\}, a < b$; and the last $\frac{z^2-z}{2}$ elements of $\mathrm{cvec}(\mv{X}) $ correspond to the (scaled) imaginary part of the lower off-diagonal elements of $\mv{X}$, i.e.,  $j\frac{[\mv{X}]_{ab} - [\mv{X}]_{ba}}{\sqrt{2}}$'s, $\forall a,b\in\{1,\ldots,z\}, a < b$.} Therefore, by defining $\hat{\mv g} = \mathrm{cvec}(\hat{\mv{G}}) \in \mathbb{R}^{M_T^2 \times 1}$ and ${\mv s}_n^{(1)} = \mathrm{cvec}(\mv{S}_n^{(1)})\in \mathbb{R}^{M_T^2 \times 1}$, we can rewrite the $n$ linear equality constraints in problem (\ref{eqn:accpm:B}) as
\begin{align}
\frac{M_T}{P}{\mv s}_m^{(1)T}\hat{\mv g} &= \tilde Q_m, \forall m\in\{2,\ldots,n\},\nonumber\\
(\mathrm{cvec}(\mv I))^T\hat{\mv g} &= 1,\label{eqn:4}
\end{align}
which correspond to a set of $n$ linear equations with $M_T^2$ real unknowns. Based on (\ref{eqn:4}), we discuss the solution $\tilde{\mv{G}}_n^{(1)}$ in (\ref{eqn:accpm:B}) by considering the following two cases with $n\ge M_T^2$ and $n< M_T^2$, respectively.
\begin{itemize}
  \item In the case with $n\ge M_T^2$, the number of linear equations in (\ref{eqn:4}) is no smaller than that of the real unknowns. In this case, as long as the vector ${\mv s}_m^{(1)}$'s are linearly independent such that $\mathrm{rank}({\mv K}_n) = M_T^2$ with ${\mv K}_n \triangleq \left[\mathrm{cvec}(\mv I), {\mv s}_2^{(1)}, \ldots, {\mv s}_{n}^{(1)}\right]^T \in \mathbb{R}^{n \times M_T^2}$, the solution of $\hat{\mv g}$ to (\ref{eqn:4}) is unique, and so is the optimal solution $\tilde{\mv G}_n^{(1)}$ in (\ref{eqn:accpm:B}) due to the one-to-one mapping of $\mathrm{cvec}(\cdot)$. As a result, the obtained $\tilde{\mv G}_n^{(1)}$ is indeed the exact MIMO channel $\bar{\mv G}$ to be learned, i.e., $\tilde{\mv G}_n^{(1)} = \bar{\mv G}$.
\item In the case of $n< M_T^2$, the set of equations in (\ref{eqn:4}) corresponds to an underdetermined system of linear equations. In this case, the solution of $\hat{\mv g}$ to (\ref{eqn:4}) is not unique in general, and so is the obtained $\tilde{\mv G}_n^{(1)}$ in (\ref{eqn:accpm:B}). As a result, $\tilde{\mv G}_n^{(1)}$ may not be the exact MIMO channel $\bar{\mv G}$, i.e., channel estimation error can occur.
\end{itemize}
From the above discussion, it is evident that when $B\to\infty$, a minimum number of $M_T^2$ feedback intervals are sufficient for the ET to obtain an exact estimate of the MIMO channel with the energy quantization based scheme.

\section{Energy Comparison Based Feedback and Channel Learning}\label{sec:comparison:feedback}

In this section, we propose an alternative feedback design so-called energy comparison, where the $B$ feedback bits per interval indicate the increase or decrease of the harvested energy at the ER in the current interval as compared to those in the previous $B$ intervals, respectively. For this scheme, we denote the feedback at the ER as $\left\{f_n^{(2)}\right\}$, the transmit covariance matrices at the ET as $\left\{\mv S_n^{(2)}\right\}$, the updated sequence of analytic centers as $\left\{\tilde{\mv G}_n^{(2)}\right\}$, and the estimated MIMO channel as $\tilde{\mv G}^{(2)}$.

\subsection{Feedback Design at ER}

In the energy comparison based scheme, the ER designs each of the $B$ bits in $f_n^{(2)}$ separately. Consider a particular feedback interval $n\in\{2,\ldots,N\}$ and let the $b$th bit of $f_n^{(2)}$ be denoted by $f^{(2)}_{n,b}$, $b\in\{1,\ldots,B\}$. Then we use $f^{(2)}_{n,b}$ to indicate the increase or decrease of the normalized harvested energy $\bar Q_n$ by the ER in the current interval $n$ as compared to $\bar Q_{n-b}$ in the $b$th previous interval, $b\in\{1,\ldots,B\}$. More explicitly, we have
\begin{eqnarray} \label{eqn:OneBitFeedback}
f_{n,b}^{(2)}=\left\{\begin{array}{ll} 1, &
{\rm if~} \bar Q_n < \bar Q_{n-b} \\ -1, & {\rm if~} \bar Q_n \ge \bar Q_{n-b} \end{array}\right..
\end{eqnarray}
Here, we assume $\bar Q_{n-b} = 0$ and $ \mv S_{n-b}^{(2)} = \mv 0$ for any $n$ and $b$ with $n-b\le 0$ for notational convenience. By combining the obtained $\left\{f_{n,b}^{(2)}\right\}_{b=1}^B$ as $f_n^{(2)}$, the ER sends it back to the ET.

After receiving the feedback $f_n^{(2)}$, at each interval $n\in\{2,\ldots,N\}$, the ET extracts the following $\min(B,n-1)$ linear inequalities on the MIMO channel $\bar{\mv G}$ based on (\ref{eqn:OneBitFeedback}) as well as (\ref{eqn:barQ}).{\footnote{Note that for any $b \ge n$ in (\ref{eqn:ComFB}), $f_{n,b}^{(2)} = -1$ always holds due to $\mv S_{n-b}^{(2)} = \mv 0$. In this case, the inequality due to $f_{n,b}^{(2)} = -1$ becomes $\mathrm{tr}\left({ \bar{\mv G}}\mv S_{n}^{(2)}\right) \ge 0$, which is trivial and does not provide any information on the MIMO channel $\bar{\mv G}\succeq \mv 0$ (due to $\mv S_{n}^{(2)}\succeq \mv 0$). Therefore, only $f_{n,b}^{(2)}$'s with $b \le \min(B,n-1)$ are used for generating inequalities for channel learning.}}
\begin{align}\label{eqn:ComFB}
f_{n,b}^{(2)}\mathrm{tr}\left({ \bar{\mv G}}\left(\mv S_{n}^{(2)} - \mv S_{n-b}^{(2)}\right)\right) \le 0, b\in\{1,\ldots,\min(B,n-1)\},
\end{align}
which correspond to $C_n=\min(B,n-1)$ cutting planes in the general expression (\ref{eqn:general:cut}) to localize the MIMO channel with $\mv \Sigma_{n,b} = f_{n,b}^{(2)}\left(\mv S_{n}^{(2)} - \mv S_{n-b}^{(2)}\right)$ and $\gamma_{n,b} = 0, b\in\{1,\ldots,\min(B,n-1)\}$.

\subsection{Training Signal Design and Channel Estimation at ET}

Based on the feedback $f_n^{(2)}$'s, the ET can obtain the working set for the interval $n\in\{1,\ldots,N\}$ similarly as in (\ref{eqn:working:set})  by replacing the cutting planes in (\ref{eqn:general:cut}) as those in (\ref{eqn:ComFB}):
 \begin{align}
\mathcal P_n^{(2)} = &\bigg\{\hat{\mv G}\big|\hat{\mv G} \succeq \mv 0, \mathrm{tr}\left(\hat {\mv G}\right)=1, \nonumber\\
&f_{m,b}^{(2)}\mathrm{tr}\left({ \hat{\mv G}}\left(\mv S_{m}^{(2)} - \mv S_{m-b}^{(2)}\right)\right) \le 0, \nonumber\\
&\forall b\in\{1,\ldots,\min(m-1,B)\}, \forall m\in\{2,\ldots,n\}\bigg\}.\label{eqn:working:set:2}
\end{align}
Accordingly, the analytic center of $\mathcal P_n^{(2)}$ is expressed as
\begin{align}
&\tilde{\mv{G}}_n^{(2)} = \mathrm{arg}\mathop\mathrm{min}\limits_{\hat{\mv G} \succeq \mv 0, \mathrm{tr}(\hat{\mv G}) = 1} -\log\det\left(\hat{\mv G}\right) \nonumber\\
&~~-\sum_{m=2}^n\sum_{b=1}^{\min(m-1,B)}\log\left(-f_{m,b}^{(2)}\mathrm{tr}\left( \hat{\mv G}\left(\mv S_{m}^{(2)} - \mv S_{m-b}^{(2)}\right)
\right)\right).\label{eqn:compar}
\end{align}

Finally, we design the transmit covariance matrix $\mv S_n^{(2)}$ for the interval $n \in \{2,\ldots,N\}$ to make the cutting planes in (\ref{eqn:ComFB}) neutral with respect to $\tilde{\mv G}_{n-1}^{(2)}$ (thus, satisfying (\ref{eqn:cut:general}) for ACCPM), i.e.,
\begin{align}\label{eqn:neutral}
\mathrm{tr}\left( \tilde{\mv{G}}_{n-1}^{(2)}\left(\mv S_{n}^{(2)} - \mv S_{n-b}^{(2)}\right)\right) = 0, b\in\{1,\ldots,\min(n-1,B)\},
\end{align}
where the above equalities hold regardless of the sign of $f_{n,b}^{(2)}$. However, if $\min(n-1,B) > 1$, then it is infeasible to find such an $\mv S_n^{(2)}$ to satisfy all these equations in (\ref{eqn:neutral}) at the same time. To overcome this difficulty, we propose to only ensure the neutral cutting plane for $b=1$, i.e., $\mathrm{tr}\left( \tilde{\mv{G}}_{n-1}^{(2)}\left(\mv S_{n} ^{(2)}- \mv S_{n-1}^{(2)}\right)\right) = 0$. Towards this end, we set
\begin{align}\label{eqn:Com:C_n}
\mv S_{n} ^{(2)}= \mv S_{n-1}^{(2)} + \mv {\Delta}_{n}.
\end{align}
Here, $\mv {\Delta}_{n}$ is a Hermitian probing matrix, which is neither positive nor negative semi-definite in general and satisfies
\begin{align}\label{eqn:delta}
\mathrm{tr}\left( \tilde{\mv{G}}_{n-1}^{(2)}\mv \Delta_n\right) = 0, n = \{2,\ldots,N\}.
\end{align}
Accordingly, the $\min(n-1,B)$ cutting planes in (\ref{eqn:ComFB}) can be re-expressed as
\begin{align}
 f_{n,b}^{(2)}\mathrm{tr}\big(\bar{ \mv G}(\mv \Delta_{n}+\cdots &+ \mv \Delta_{n-b+1})\big) \le 0,\nonumber\\
 & b\in\{1,\ldots,\min(n-1,B)\}.\label{eqn:cuttingplanes:comparison}
  \end{align}
To find such a ${\mv{\Delta}}_n$ to satisfy (\ref{eqn:delta}) for the $n$th interval, we express $\tilde{\mv{g}}_{n-1}^{(2)} = \mathrm{cvec}\left(\tilde{\mv{G}}_{n-1}^{(2)}\right)$ and ${\mv{\delta}}_n=\mathrm{cvec}\left({\mv{\Delta}}_n\right)$, where $\tilde{\mv{g}}^{(2)T}_{n-1}{\mv{\delta}}_n=\mathrm{tr}\left(\tilde{\mv{G}}_{n-1}^{(2)}{\mv{\Delta}}_n\right)=0$. Due to the one-to-one mapping of $\mathrm{cvec}\left(\cdot\right)$, finding ${\mv{\Delta}}_n$ is equivalent to finding ${\mv{\delta}}_n$ that is orthogonal to $\tilde{\mv{g}}_{n-1}^{(2)}$. Define a projection matrix $\mv{F}_n = \mv{I} - \frac{\tilde{\mv{g}}_{n-1}^{(2)}\tilde{\mv{g}}_{n-1}^{(2)T}}{\|\tilde{\mv{g}}_{n-1}^{(2)}\|^2}$. Then we can express  ${\mv{F}}_n = {\mv{V}}_n{\mv{V}}^T_n$, where ${\mv{V}}_n \in \mathbb{R}^{M_T^2\times (M_T^2-1)}$ satisfies ${\mv{V}}^T_n\tilde{\mv{g}}_{n-1}^{(2)}= \mv{0}$ and ${\mv{V}}^T_n{\mv{V}}_n = \mv{I}$. Thus, $\mv{\delta}_n$ can be any vector in the subspace spanned by ${\mv{V}}_n$. Specifically, we set
\begin{align}\label{learning:3}
\mv{\delta}_n = {\mv{V}}_n\mv{p},
\end{align}
where $\mv{p}\in\mathbb{R}^{(M_T^2-1)\times 1}$ is a randomly generated CSCG vector with zero-mean and identity covariance matrix (generally scaled by a certain positive constant; see footnote 12) in order to make $\mv{\delta}_n$ independently drawn from the subspace. With the obtained $\mv{\delta}_n$, we have $\mv{\Delta}_n = \mathrm{cmat}(\mv{\delta}_n)$,{\footnote{\color{black}Note that $\mv{\Delta}_n$ in general contains both positive and negative eigenvalues. As a result, the update in (\ref{eqn:Com:C_n}) may not necessarily yield an $\mv{S}_n$ that satisfies both $\mathrm{tr}(\mv{S}_n)\leq P$ and $\mv{S}_n\succeq \mv{0}$. Nevertheless, by setting $\|\mv{p}\|$ to be sufficiently smaller than $P$, we can always find a $\mv{p}$ and its resulting $\mv{S}_n$ satisfying the above two conditions with only a few random trials. In this paper, we choose  $\|\mv{p}\|=P/5$.}} where $\mathrm{cmat}(\cdot)$ denotes the inverse operation of $\mathrm{cvec}(\cdot)$. Accordingly, $\mv{S}_n$ that satisfies the equation in (\ref{eqn:neutral}) for $b=1$ is obtained.

In summary, the energy comparison based feedback and channel learning scheme is described as Algorithm 3 in Table \ref{table:3}.

\begin{table}[!t]\scriptsize
\caption{Energy Comparison Based Scheme}
\label{table:3} \centering
\begin{tabular}{|p{8.5cm}|}
\hline
\textbf{Algorithm 3}\\
\hline\vspace{0.01cm}
  1) {\bf Initialization:} Set $n=0$, and $\mv{S}_1^{(2)}={P}/{M_T}\mv{I}$. \\
2) {\bf Repeat:}
  \begin{itemize} \setlength{\itemsep}{0pt}
    \item[a)] $n \gets n+1$;
    \item[b)] The ET transmits with $\mv{S}_n^{(2)}$;
    \item[c)] The ER sends back $\{f_{n,b}^{(2)}\}_{b=1}^{B}$ (if $n\ge 2$) to the ET with $f^{(2)}_{n,b}=-1$ (or $1$) if $\bar Q_n \ge \bar Q_{n-b}$ (otherwise), $b\in\{1,\ldots,B\}$;
    \item[d)] The ET updates the working set as $\mathcal P_n^{(2)}$ using (\ref{eqn:working:set:2}), and computes the analytic center of $\mathcal P_n^{(2)}$ as $\tilde{\mv G}_{n}^{(2)}$ given in (\ref{eqn:compar});
    \item[e)] The ET generates $\mv{\Delta}_{n+1}$, and updates $\mv{S}_{n+1}^{(2)}=\mv{S}^{(2)}_{n}+\mv{\Delta}_{n+1}$.
  \end{itemize}
  3) {\bf Until}  $n\ge N$.\\
  4) The ET estimates the MIMO channel as $\tilde{\mv{G}}^{(2)}={\tilde{\mv{G}}^{(2)}_{N}}$.\\
 \hline
\end{tabular}\vspace{0em}
\end{table}	

\subsection{Energy Quantization Versus Energy Comparison}\label{sec:comparsion2}

\begin{figure}
\centering
 \epsfxsize=1\linewidth
    \includegraphics[width=8cm]{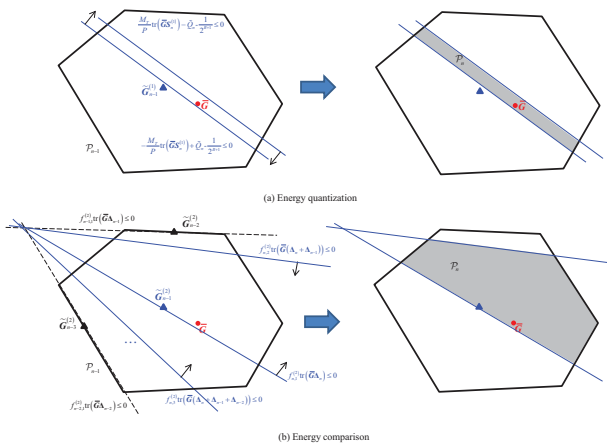}
\caption{Energy quantization versus energy comparison based schemes: the case with large $B$.} \label{fig:comparison:1}
\end{figure}

\begin{figure}
\centering
 \epsfxsize=1\linewidth
    \includegraphics[width=8cm]{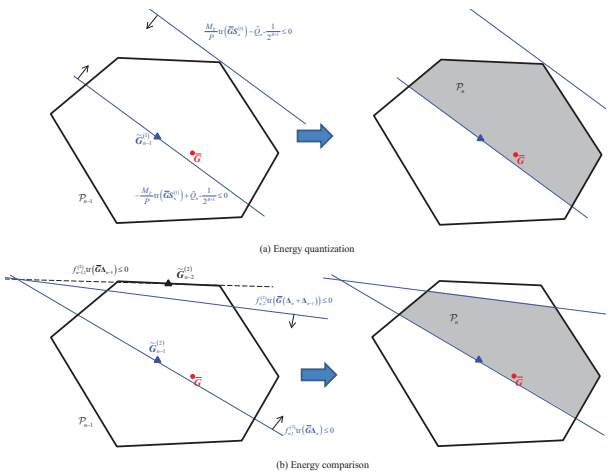}
\caption{Energy quantization versus energy comparison based schemes: the case with small $B$.} \label{fig:comparison:2}
\end{figure}

In this subsection, we compare the energy quantization versus energy comparison based feedback and channel learning schemes via geometric interpretations on their cutting planes at each feedback interval $n \in \{2,\ldots,N\}$. For the ease of comparison, we assume that both schemes have the same working set $\mathcal P_{n-1} = \mathcal P_{n-1}^{(1)}= \mathcal P_{n-1}^{(2)}$ and the same analytic center $\tilde{\mv G}_{n-1}^{(1)} = \tilde{\mv G}_{n-1}^{(2)}$ at interval $n-1$, as shown in Figs. \ref{fig:comparison:1} and \ref{fig:comparison:2}, for the cases of large $B$ and small $B$ values, respectively.

First, we compare graphically the cutting planes returned by the two schemes. For the case of energy quantization, there are two cutting planes obtained at each interval $n$ as specified in (\ref{eqn:inequations:2:1}) and (\ref{eqn:inequations:2}). It is observed in  Figs. \ref{fig:comparison:1}(a) and \ref{fig:comparison:2}(a) that the two cutting planes are parallel to each other, since the gradient of the left-hand-side (LHS) in (\ref{eqn:inequations:2:1}) is just the opposite of that in (\ref{eqn:inequations:2}). It is also observed that only one of the two cutting planes is deep (e.g., see Fig. \ref{fig:comparison:1}(a)) or neutral (e.g., see Fig. \ref{fig:comparison:2}(a)) with respect to the analytic center $\tilde{\mv G}^{(1)}_{n-1}$ of $\mathcal P_{n-1}$, while the other is neither deep nor neutral. This is consistent with the training signal design in (\ref{neutral:1}).

In contrast, for the case of energy comparison, there are a total of $\min(n-1,B)$ cutting planes in each interval $n$ as specified in (\ref{eqn:cuttingplanes:comparison}). It is observed in Figs. \ref{fig:comparison:1}(b) and \ref{fig:comparison:2}(b) that all these cutting planes pass through the common point of $\mv 0$, while among them, only the neutral cutting plane corresponding to $b=1$ (i.e., $f_{n,1}^{(2)} \mathrm{tr}(\bar{\mv G}\mv \Delta_{n}) \le 0$) passes through the analytic center $\tilde{\mv G}_{n-1}^{(2)}$ of $\mathcal P_{n-1}$. Furthermore, the non-neutral cutting plane with $b=2$ (i.e., $f_{n,2}^{(2)} \mathrm{tr}(\bar{\mv G}(\mv \Delta_{n}+\mv \Delta_{n-1})) \le 0$) is observed to locate between the neutral cutting plane in the current interval $n$ with $b=1$ (i.e., $f_{n,1}^{(2)} \mathrm{tr}(\bar{\mv G}\mv \Delta_{n}) \le 0$) and that in the previous interval $n-1$ with $b=1$ (i.e., $f_{n-1,1}^{(2)} \mathrm{tr}(\bar{\mv G}\mv \Delta_{n-1}) \le 0$, which passes through the analytic center $\tilde{\mv G}_{n-2}^{(2)}$ of $\mathcal P_{n-2}$), since the gradient of the LHS in the former inequality is a linear combination of those in the latter two. Note that the point $\tilde{\mv G}_{n-2}^{(2)}$ always locates on the boundary of $\mathcal P_{n-1}$; as a result, the cutting plane with $b = 2$ is very likely to cross the interior of $\mathcal P_{n-1}$, making it efficiently eliminate a sizable portion of $\mathcal P_{n-1}$. Similar properties also hold for other non-neutral cutting planes with $b \in \{3,\ldots,\min(n-1,B)\}$.

Next, with the above geometric interpretations, we compare the effectiveness of the two feedback schemes for channel learning by considering three cases with different values of $B$.
\begin{itemize}
  \item First, consider the case of large $B$ as shown in Fig. \ref{fig:comparison:1}. For energy quantization, the gap between the two cutting planes is small (particularly, it is zero when $B\to \infty$), and as a result they can cut a significant portion of $\mathcal P_{n-1}$. Therefore, energy quantization is more effective than energy comparison in this case.
  \item Next, consider the case of small $B$ (but with $B>1$) as depicted in Fig. \ref{fig:comparison:2}. For energy quantization, the gap between the two cutting planes becomes significantly larger as compared to the case of large $B$ in Fig. \ref{fig:comparison:1}. As a result, the non-neutral cutting plane is likely to be less efficient or even redundant. In contrast, for energy comparison, the non-neutral cutting planes will still cross the interior of $\mathcal P_{n-1}$, which helps cut the volume of $\mathcal P_{n-1}$ effectively even with a small $B$ value. As a result, the cutting planes of energy comparison are generally more effective than those of energy quantization in this case.
  \item Finally, consider the special case of $B=1$. In this case, both schemes can only return one valid cutting plane per interval. Intuitively, the two schemes should have similar channel learning performance; however, as will be shown by the numerical results in Section \ref{sec:numerical:results}, energy comparison is still more effective than energy quantization in this case.
\end{itemize}

\section{Complexity Reduction Via Pruning Irrelevant Cutting Planes}\label{sec:low:complexity}

In the previous sections, we have proposed a general channel learning design framework based on ACCPM as well as two specific feedback schemes for the ET to learn the CSI, where at each feedback interval the ET needs to solve an optimization problem (see (\ref{eqn:accpm:general})) for obtaining the analytic center of the updated working set. As the number of feedback intervals $n$ increases and that of added cutting planes becomes large, the complexity for solving such problems also increases significantly. To solve this issue, in this section we propose an efficient complexity reduction method by pruning irrelevant cutting planes at each feedback interval $n$ after obtaining the analytic center of $\mathcal P_n$  (i.e., by adding a pruning procedure between steps 2-d) and 2-e) in Algorithms 1-3). For brevity, we discuss for the general design framework in Section \ref{sec:Framework} only, while similar procedures can be applied for the two specific schemes in Sections \ref{sec:quantization:feedback} and \ref{sec:comparison:feedback}.

\begin{figure}
\centering
 \epsfxsize=1\linewidth
    \includegraphics[width=5cm]{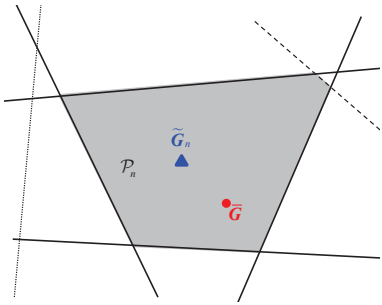}
\caption{An example working set $\mathcal P_n$ (shown shaded) characterized by various cutting planes (depicted as lines). Here, the solid lines represent most important cutting planes for bounding $\mathcal P_n$; the dashed line shows a less important cutting plane (though still active for bounding $\mathcal P_n$); and the dotted line is a redundant cutting plane.} \label{fig:pruning}
\end{figure}

The complexity reduction method is motivated by the example illustrated in Fig. \ref{fig:pruning}, where various cutting planes returned at intervals $2,\ldots,n$ establish a working set $\mathcal{P}_n$ for localizing the target set $\varphi = \left\{\bar{\mv G}\right\}$. It is observed that among these cutting planes, some of them are important active constraints to bound $\mathcal{P}_n$ (see the solid lines in Fig. \ref{fig:pruning}), while the others are less relevant or even redundant (see the dashed and dotted lines in Fig. \ref{fig:pruning}). Therefore, by pruning such irrelevant cutting planes, we could reduce the computation complexity for solving problem (\ref{eqn:accpm:general}) without degrading the channel learning performance. In the following, we first rank the irrelevance of the cutting planes.

Consider a particular feedback interval $n\in\{2,\ldots,N\}$ and suppose that there are $\sum_{m=2}^nC_m$ cutting planes at hand, which form a polyhedron as
\begin{align}
\bigg\{\hat{\mv G}\big|&\mathrm{tr}(\mv \Sigma_{m,c} \hat{\mv G}) - \gamma_{m,c}  \le 0, \nonumber\\& \forall c\in\{1,\ldots,C_m\}, \forall m\in\{2,\ldots,n\}\bigg\}.\label{eqn:polyhedron}
\end{align}
As compared to the working set $\mathcal P_n$ in (\ref{eqn:working:set}), we have omitted $\hat{\mv G} \succeq \mv 0$ and $\mathrm{tr}(\hat{\mv G}) = 1$ in (\ref{eqn:polyhedron}) to focus on studying the cutting planes only. For simplicity, we use $\tilde{\mv G}_{n}$ (i.e., the analytic center of $\mathcal P_n$ in (\ref{eqn:working:set})) as the (approximate) analytic center of the polyhedron in (\ref{eqn:polyhedron}), since the constraints $\hat{\mv G} \succeq \mv 0$ and $\mathrm{tr}(\hat{\mv G}) = 1$ become less likely to be active when the number of cutting planes $\sum_{m=2}^nC_m$ becomes large. Define
\begin{align}
&\mv \Psi_n(\hat{\mv G}) = \nonumber\\& \sum_{m=2}^{n}\sum_{c=1}^{C_m}  \left(\mathrm{tr}\left(\mv \Sigma_{m,c} \hat{\mv G}\right) - \gamma_{m,c}\right)^{-2}\mathrm{cvec}(\mv \Sigma_{m,c})(\mathrm{cvec}(\mv \Sigma_{m,c}))^T.
\end{align}
Then, the ellipsoid
\begin{align*}
\xi_{\rm in} = \bigg\{\hat{\mv G}\bigg|& \left(\mathrm{cvec}\left(\hat{\mv G}-\tilde{\mv G}_n\right)\right)^T \mv \Psi_n(\tilde{\mv G}_n)\nonumber\\&\left(\mathrm{cvec}\left(\hat{\mv G}-\tilde{\mv G}_n\right)\right) \le 1\bigg\}
\end{align*}
lies inside the polyhedron in (\ref{eqn:polyhedron}), while the ellipsoid
\begin{align*}
\xi_{\rm out} = \bigg\{\hat{\mv G}\bigg|& \left(\mathrm{cvec}\left(\hat{\mv G}-\tilde{\mv G}_n\right)\right)^T \mv \Psi_n(\tilde{\mv G}_n)\nonumber\\
&\left(\mathrm{cvec}\left(\hat{\mv G}-\tilde{\mv G}_n\right)\right) \le \left(\sum\nolimits_{m=2}^nC_m\right)^2\bigg\},
\end{align*}
which is $\xi_{\rm in}$ scaled by a factor of $\sum_{m=2}^nC_m$ over its center, contains the polyhedron in (\ref{eqn:polyhedron}) \cite[Chapter 8.5]{BV:Convex}. Therefore, the ellipsoid $\xi_{\rm in}$ at least grossly approximates the shape of the polyhedron in (\ref{eqn:polyhedron}), and indicates the following irrelevance measure \cite{ACCPM}
\begin{align}\label{eqn:measure:irrelevance}
\eta_{m,c} &= \frac{-\mathrm{tr}(\mv \Sigma_{m,c} \tilde{\mv G}_{n}) + \gamma_{m,c} }{\sqrt{(\mathrm{cvec}(\mv \Sigma_{m,c} ))^T\mv \Psi_n^{-1}(\tilde{\mv G}_n)\mathrm{cvec}(\mv \Sigma_{m,c} )}}
\end{align}
for the cutting plane $\mathrm{tr}(\mv \Sigma_{m,c} \hat{\mv G}) - \gamma_{m,c}  \le 0,  c\in\{1,\ldots,C_m\},m\in\{2,\ldots,n\}$. Here, larger $\eta_{m,c}$ means that the corresponding cutting plane is more likely to be irrelevant.

Next, we prune irrelevant cutting planes based on the above irrelevance measure $\{\eta_{m,c}\}$. We simply drop the cutting planes with largest values of $\eta_{m,c}$, and keep the cutting planes with smaller values. In particular, the number of kept cutting planes is fixed as $N_c > 0$, where $N_c$ is a design parameter controlling the trade-off between the channel learning accuracy and computation complexity (as will be shown in more detail via numerical results in Section \ref{sec:pruning}). With fixed $N_c$, the computational complexity per interval for solving problem (\ref{eqn:accpm:general}) does not grow as $n$ increases.

\section{Numerical Results}\label{sec:numerical:results}

In this section, we provide numerical results to evaluate the performance of our proposed channel learning schemes. We consider a point-to-point MIMO WET system with one ER located at a distance of 5 meters from the ET, where the average path loss from the ET to the ER is 40 dB. For the considered short transmission distance, the LOS signal is dominant, and thus the Rician fading is used to model the channel from the ET to the ER. Specifically, we assume
\begin{align}
\mv{H}= \sqrt{\frac{\chi_R}{1+\chi_R}}\mv{H}^{\rm{LOS}}+\sqrt{\frac{1}{1+\chi_R}}\mv{H}^{\rm{NLOS}},
\end{align}
where $\mv{H}^{\rm{LOS}}\in\mathbb{C}^{M_R\times M_T}$ is the LOS deterministic component, $\mv{H}^{\rm{NLOS}} \in\mathbb{C}^{M_R\times M_T}$ denotes the non-LOS Rayleigh fading component with each element being an independent CSCG random variable with zero mean and variance of $10^{-4}$ (to be consistent with the assumed average power attenuation of $40$ dB), and $\chi_R$ is the Rician factor set to be $5$ dB. For the LOS component, we use the far-field uniform linear antenna array model with each row of $\mv{H}^{\rm{LOS}}$ expressed as $10^{-2}\left[1~ e^{j\theta}~\cdots~e^{j(M_T-1)\theta}\right]$ with $\theta = -\frac{2\pi \kappa\sin(\phi)}{\lambda}$, where  $\kappa$ is the spacing between two adjacent antenna elements at the ET, $\lambda$ is the carrier wavelength, and $\phi$ is the direction of the ER from the ET. We set $\kappa=\frac{\lambda}{2}$ and $\phi = 30^\circ$. Furthermore, we set the number of transmit antennas at the ET as $M_T=4$, the number of receive antennas at the ER as $M_R = 2$, and the maximum transmit sum-power as $P = 30$ dBm ($1$ Watt).

\subsection{Channel Learning Performance}\label{sec:numerical:channelLearning}

\begin{figure}
\centering
 \epsfxsize=1\linewidth
    \includegraphics[width=8cm]{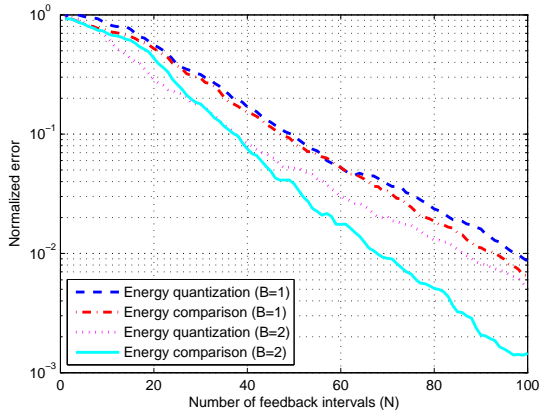}
\caption{Normalized error of estimated matrix norm versus the number of feedback intervals $N$ with $B=1$ and $B=2$.}
\label{fig:ChannelLearning1}
\end{figure}

\begin{figure}
\centering
 \epsfxsize=1\linewidth
    \includegraphics[width=8cm]{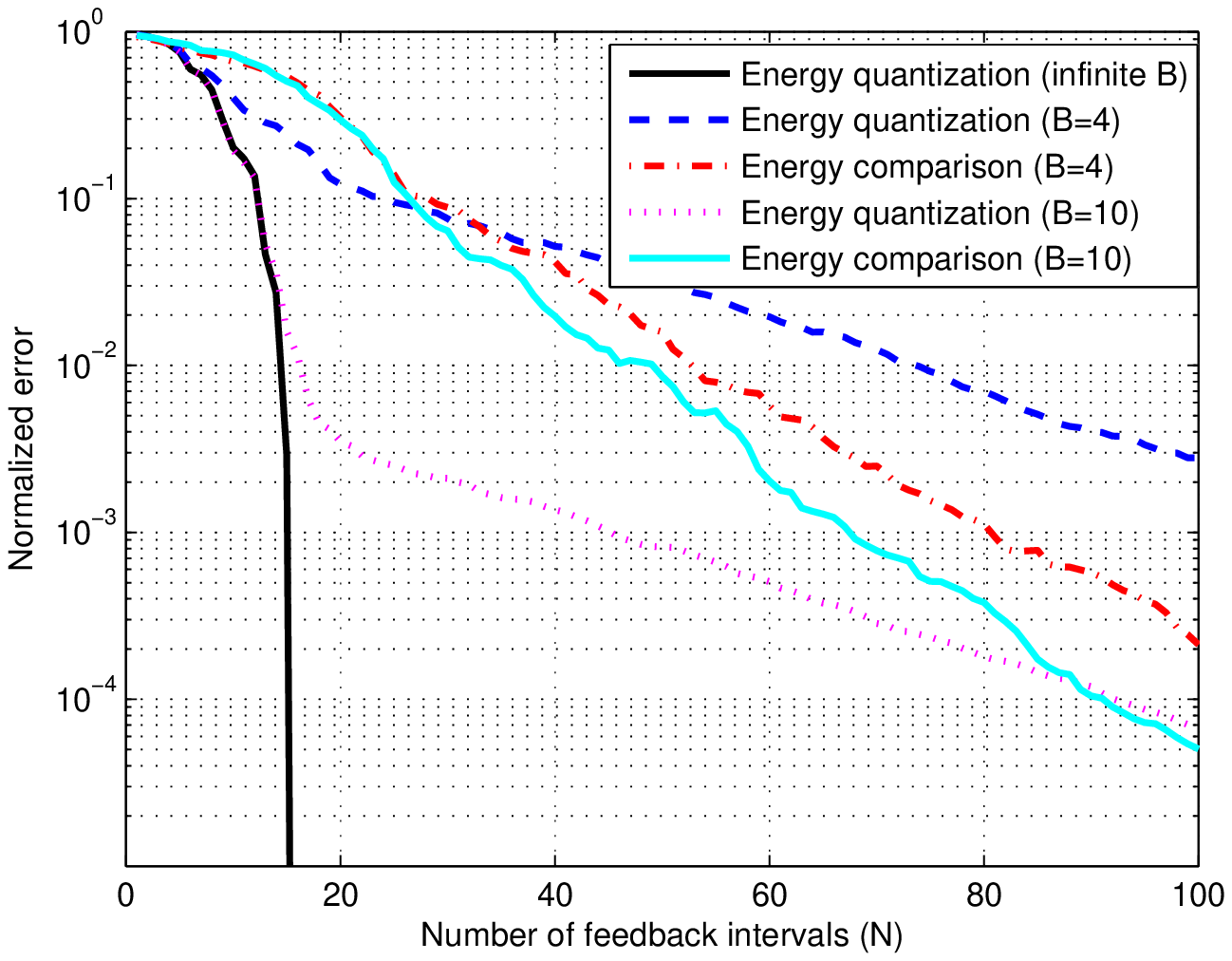}
\caption{Normalized error of estimated matrix norm versus the number of feedback intervals $N$ with $B=4$, $B=10$, and $B\to\infty$.}
\label{fig:ChannelLearning2}
\end{figure}

To start with, we compare the channel learning performance of the energy quantization and energy comparison based feedback schemes. Figs. \ref{fig:ChannelLearning1} and \ref{fig:ChannelLearning2} plot the normalized error of estimated matrix norm, i.e., $\left\|\tilde{\mv G} - \bar{\mv G}\right\|_{\rm F}$, versus the number of feedback intervals $N$, under different numbers of feedback bits per interval, $B$. When $B=1$, it is observed in Fig. \ref{fig:ChannelLearning1} that energy comparison achieves a lower normalized error than energy quantization. When $B>1$, it is observed in Figs. \ref{fig:ChannelLearning1} and \ref{fig:ChannelLearning2} that energy quantization outperforms energy comparison when $N$ is below a certain threshold (e.g., $N \le 30$ for $B=2$, and $N\le 90$ for $B=10$), while the opposite is true when $N$ becomes larger than the threshold. This result can be explained as follows based on the geometric interpretations of their different cutting planes (for a given interval $n$) in Section \ref{sec:comparsion2}. When the number of feedback intervals $n$ is small, the working set $\mathcal P_n$ is large and energy quantization is more effective in space cutting with a fixed $B$; however, when $n$ is large, $\mathcal P_n$ becomes small and energy comparison is more effective due to multiple cutting planes with the same $B$ value. Furthermore, when $B\to \infty$, it is observed in Fig. \ref{fig:ChannelLearning2} that energy quantization obtains an exact estimate of the MIMO channel with the normalized error falling sharply to zero after $N\ge M_T^2 = 16$ intervals. This result is consistent with our discussion in Section \ref{sec:perfect:feedback}.

\subsection{Energy Beamforming Gain}\label{sec:numerical:EBGain}

\begin{figure}
\centering
 \epsfxsize=1\linewidth
    \includegraphics[width=8cm]{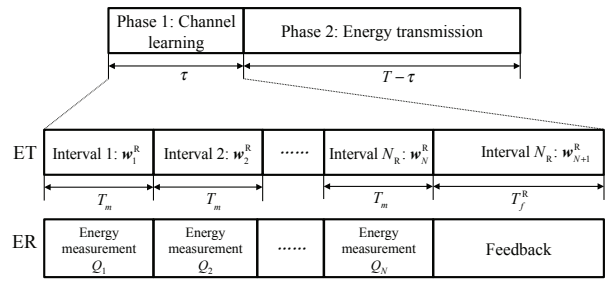}
\caption{The two-phase transmission protocol for MIMO WET with random beamforming.}
\label{fig:twophase:ref}
\end{figure}

\begin{figure}
\centering
 \epsfxsize=1\linewidth
    \includegraphics[width=8cm]{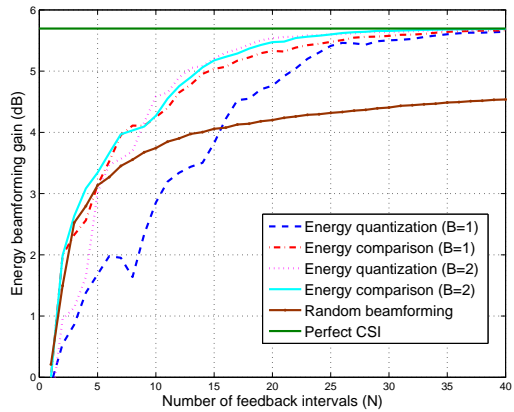}
\caption{Energy beamforming gain versus the number of feedback intervals $N$ with $B=1$ and $B=2$.}
\label{fig:EnergBeamGain1}
\end{figure}

\begin{figure}
\centering
 \epsfxsize=1\linewidth
    \includegraphics[width=8cm]{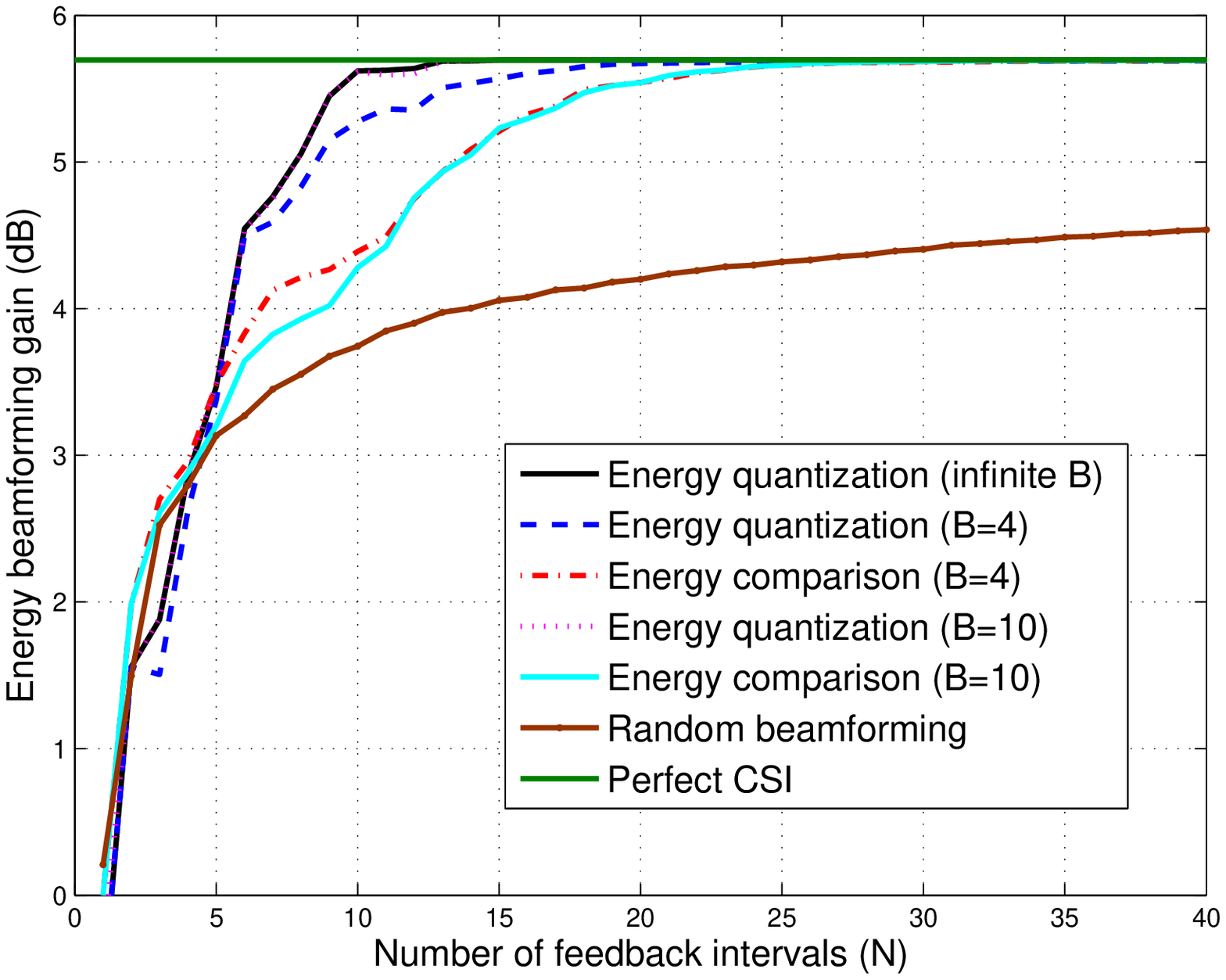}
\caption{Energy beamforming gain versus the number of feedback intervals $N$ with $B=4$, $B=10$, and $B\to\infty$.}
\label{fig:EnergBeamGain2}
\end{figure}

Next, we evaluate the energy beamforming performance with estimated MIMO channels by the two schemes, as compared to a reference scheme named random beamforming. Similar to our proposed ACCPM based channel learning, the random beamforming scheme implements a two-phase transmission protocol for channel learning and energy transmission, respectively,  as shown in Fig. \ref{fig:twophase:ref}. However, the channel learning phase in this case is divided into $N_R+1$ intervals, with the length of $T_m$ for the first $N_R$ intervals and $T_f^{\rm R}$ for the last one. In the $n$th interval, $n\in\{1,\ldots,N_R\}$, the ET randomly generates a beamforming vector $\mv{w}_n^{\rm R}$ with $\left\|\mv{w}_n^{\rm R}\right\|^2 = P$ and transmits with the covariance matrix $\mv S_n^{\rm R} = \mv{w}_n^{\rm R}\mv{w}_n^{{\rm R}H}$. At the same time, the ER measures the harvested energy $Q_n^{\rm R} = T_m\mathrm{tr}(\mv G\mv S_n^{\rm R})$. In the $(N_R+1)$th interval, the ER compares the measured energy amounts  $\left\{Q_n^{\rm R}\right\}_{n=1}^{N_R}$ to obtain the maximum value, and sends its index $n_{\max} = \arg\max_{n\in\{1,\ldots,N_R\}} Q_n^{\rm R}$ back to the ET. Hence, the number of feedback bits in the $(N_R+1)$th interval is given by $B_{\rm R} =\lceil \log_2(N_R)\rceil$. After receiving the feedback information, in the second energy transmission phase, the ET transmits with the energy beamforming vector $\mv w_{n_{\max}}^{\rm R}$ in the $n_{\max}$th interval of the channel learning phase. For a fair comparison, we set the length of the channel learning phase in the random beamforming scheme to be same as that in our proposed ACCPM based channel learning scheme, i.e., $N_RT_m + T_f^{\rm R} = N(T_m+T_f)$. Furthermore, to obtain an accurate measurement result, the time consumed is in general larger than that for the feedback in each interval; hence, we assume $T_m = 2T_f$ and $T_f = T_f^{\rm R}$ in the simulation by ignoring the time difference in sending back different number of bits. Thus, we have $N_R = \lfloor \frac{3N-1}{2}\rfloor$.

Figs. \ref{fig:EnergBeamGain1} and \ref{fig:EnergBeamGain2} depict the achieved energy beamforming gain $\tilde \chi$ given by (\ref{eqn:chiEB}) in dB scale versus the number of feedback intervals $N$ under different values of $B$. It is observed that when $N$ is small (e.g., $N<30$), energy comparison (energy quantization) achieves a higher energy beamforming gain in the case of $B=1$ ($B=4$ and $B=10$), while the two schemes have a similar energy beamforming gain when $B=2$. This result is consistent with the channel learning performance comparison in Figs. \ref{fig:ChannelLearning1} and \ref{fig:ChannelLearning2}. In addition, when $N$ becomes large (e.g., $N>40$), both schemes with different values of $B$ are observed to achieve energy beamforming gains close to the optimal one by OEB in the perfect CSI case (i.e., $\chi^*$ defined in (\ref{eqn:EBGain})). Based on this observation together with those in Figs. \ref{fig:ChannelLearning1} and \ref{fig:ChannelLearning2}, it is evident that a relatively rough channel estimation is sufficient to achieve a close-to-optimal energy beamforming gain for the MIMO WET system. Furthermore, it is observed from Figs. \ref{fig:EnergBeamGain1} and \ref{fig:EnergBeamGain2}  that the random beamforming scheme performs much worse than our proposed ACCPM based channel learning schemes. Also note that unlike our proposed algorithms, the random beamforming scheme can only estimate the optimal beamforming vector, but cannot be used to estimate the exact MIMO channel.

\begin{figure}
\centering
 \epsfxsize=1\linewidth
    \includegraphics[width=8cm]{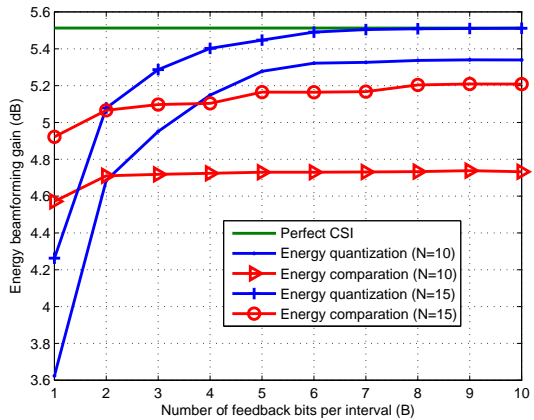}
\caption{Energy beamforming gain versus the number of feedback bits per interval $B$, under different values of $N$.}
\label{fig:EnergBeamGain3}
\end{figure}

Furthermore, it is interesting to consider a practical scenario where the number of feedback intervals $N$ is kept small. Fig. \ref{fig:EnergBeamGain3} shows the energy beamforming gain versus the number of feedback bits per interval $B$, with $N = 10$ and $N=15$. It is observed that as $B$ increases, the energy beamforming gain by energy quantization improves significantly, while that by energy comparison almost remains unchanged. This shows that energy quantization is more effective in the regime of large $B$ but limited $N$.

To summarize, the above performance comparison provides useful insights on the selection between the energy quantization and energy comparison schemes in practice. For MIMO WET systems requiring relatively coarse channel estimation, energy quantization is more effective when the number of feedback bits per interval $B$ is large, while energy comparison is more suitable when $B$ is small. However, for other applications that require more accurate channel estimation (e.g., communication systems such as cognitive radio networks \cite{NoamGoldsmith2013,GopalakrishnanSidiropoulos2014}), energy comparison could be more effective even under large values of $B$, provided that the number of feedback intervals $N$ is sufficiently large.

\subsection{Performance with Cutting Planes Pruning}\label{sec:pruning}

Furthermore, we evaluate the performance of the complexity reduction method via cutting planes pruning, in terms of both normalized error of estimated matrix norm and achieved energy beamforming gain. Due to space limitation, we only consider the energy quantization based scheme, while similar results have been observed for energy comparison and thus are omitted. We set $B=2$ in this subsection.

\begin{figure}
\centering
 \epsfxsize=1\linewidth
    \includegraphics[width=8cm]{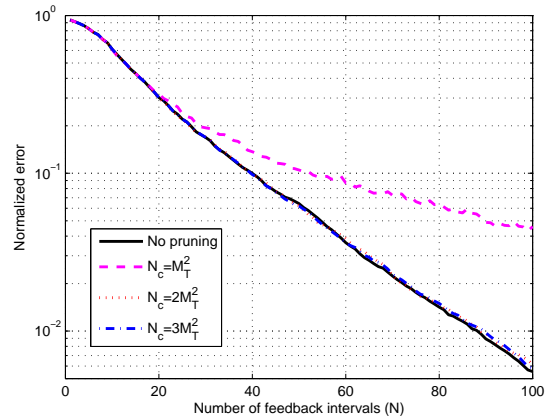}
\caption{Normalized error of estimated matrix norm versus the number of feedback intervals $N$ for energy quantization with cutting planes pruning.}
\label{fig:pruning1}
\end{figure}

Fig. \ref{fig:pruning1} shows the normalized error of estimated matrix norm versus the number of feedback intervals $N$ under different number of kept cutting planes $N_c$. It is observed that the cases with $N_c = 2M_T^2$ and $N_c = 3M_T^2$ achieve similar normalized errors of estimated matrix norm as that without pruning, while the case with $N_c= M_T^2$ results in noticeable performance degradation.

\begin{figure}
\centering
 \epsfxsize=1\linewidth
    \includegraphics[width=8cm]{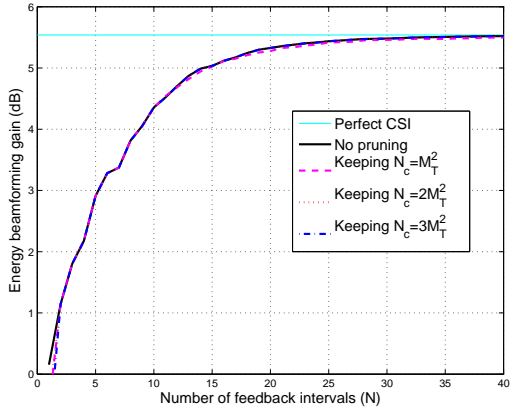}
\caption{Energy beamforming gain versus the number of feedback intervals $N$ for energy quantization with cutting planes pruning.}
\label{fig:pruning2}
\end{figure}

Fig. \ref{fig:pruning2} depicts the energy beamforming gain versus $N$. It is observed that all the cases with $N_c = M_T^2$, $N_c = 2M_T^2$, and $N_c = 3M_T^2$ achieve similar energy beamforming gains as compared to that without pruning. This demonstrates the effectiveness of the method of cutting planes pruning.

\subsection{Effect of Energy Measurement Errors}\label{sec:EnergyMeasurementError}

In the above studies, we have assumed that the energy measurement at the ER is perfect and has no errors for the purpose of exposition. In the last subsection, we investigate the effect of energy measurement errors (e.g. due to the distortion induced by the energy harvesting circuits) on the channel learning performances by simulation. It is assumed that the measured energy amount at the ER in each interval $n\in\{1,\ldots,N\}$ is $\hat Q_n = Q_n + v_n$, where $Q_n$ is the exact harvested energy given in (\ref{eqn1:measured}) and $v_n$ denotes the error in the energy meter reading which follows a uniform distribution on the interval $[-\alpha Q_n, \alpha Q_n]$ with $0\le\alpha<1$ being a constant.

Consider the general design framework for channel learning in Section \ref{sec:Framework}. Due to the energy measurement error, at each interval $n$ the cutting planes in (\ref{eqn:general:cut}) may not hold in general and thus it may occur that $\mathcal P_n \nsupseteq \varphi$. To resolve this issue, we modify the cutting planes in (\ref{eqn:general:cut}) as
\begin{align}\label{eqn:relaxed}
\mathrm{tr}(\mv \Sigma_{n,c} \bar{\mv G}) - \gamma_{n,c}  \le t_{n,c}, c\in\{1,\ldots,C_n\}, n\in\{2,\ldots,N\},
\end{align}
and accordingly obtain the relaxed working set at each interval $n\in\{1,\ldots,N\}$ as
\begin{align}
{\mathcal P}'_n = \bigg\{\hat{\mv G}\big|&\hat{\mv G} \succeq \mv 0, \mathrm{tr}\left(\hat {\mv G}\right)=1, \mathrm{tr}\left(\mv \Sigma_{m,c} \hat{\mv G}\right) - \gamma_{m,c}  \le t_{m,c}, \nonumber\\& \forall c\in\{1,\ldots,C_m\},\forall m\in\{2,\ldots,n\} \bigg\},\label{eqn:working:set:relaxed}
\end{align}
where $\{t_{n,c} \ge 0\}$ are parameters designed for ensuring ${\mathcal P}'_n \supseteq \varphi$ with improved accuracy. Specifically, we set $\{t_{n,c}\}$ as the optimal solution of $\{\hat t_{n,c}\}$ to the following convex optimization problem to minimize the distortion to the cutting planes.
\begin{align}
\min \limits_{\{\hat t_{n,c}\}, \hat{\mv G}} & ~ \sum_{m=2}^{n}\sum_{c=1}^{C_m} \hat t_{m,c} \nonumber\\
\mathrm{s.t.}&~\mathrm{tr}\left(\mv \Sigma_{m,c} \hat{\mv G}\right) - \gamma_{m,c}  \le \hat t_{m,c}, \nonumber\\&~~~~~~~~~~~~~\forall c\in\{1,\ldots,C_m\}, \forall m\in\{2,\ldots,n\}\nonumber\\
&~\hat{\mv G}\succeq \mv 0, \mathrm{tr}(\hat{\mv G}) = 1\nonumber\\
&~ \hat t_{m,c} \ge 0, \forall c\in\{1,\ldots,C_m\}, \forall m\in\{2,\ldots,n\}.\label{eqn:t_nc}
\end{align}
Then the analytic center of the modified working set ${\mathcal P}'_n$ is given by
\begin{align}
\tilde{\mv{G}}'_n = &\mathrm{arg}\mathop\mathrm{min}\limits_{\hat{\mv G}\succeq \mv 0, \mathrm{tr}(\hat{\mv G}) = 1} -\log\det\left(\hat{\mv G}\right) \nonumber\\&-\sum_{m=2}^n\sum_{c=1}^{C_m}\log\left(-\mathrm{tr}\left(\mv \Sigma_{m,c} \hat{\mv G}\right) + \gamma_{m,c} + t_{m,c}\right), \nonumber\\&~~~~~~~~~~n\in \{1,\ldots,N\}. \label{eqn:accpm:general:relaxed}
\end{align}
As a result, the modified generic algorithm can be obtained similarly as Algorithm 1 by replacing the cutting planes in (\ref{eqn:general:cut}) by those in (\ref{eqn:working:set:relaxed}), and the analytic center $\left\{\tilde{\mv{G}}_n\right\}$ by $\left\{\tilde{\mv{G}}'_n\right\}$. Note that the energy quantization and energy comparison based channel learning schemes should be accordingly modified as above in the case with energy measurement errors.

\begin{figure}
\centering
 \epsfxsize=1\linewidth
    \includegraphics[width=8cm]{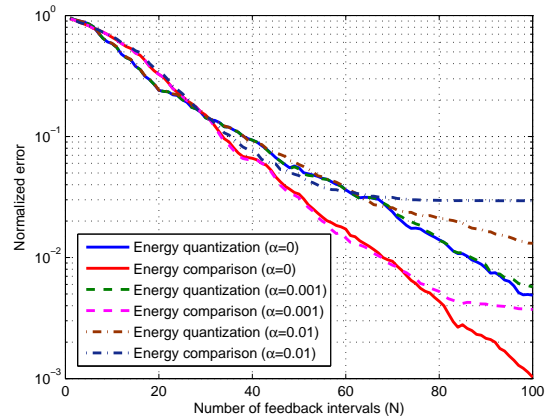}
\caption{Normalized error of estimated matrix norm versus the number of feedback intervals $N$ under energy measurement errors.}
\label{fig:error1}
\end{figure}

\begin{figure}
\centering
 \epsfxsize=1\linewidth
    \includegraphics[width=8cm]{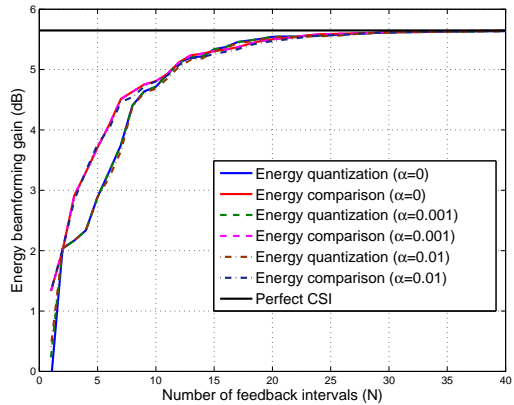}
\caption{Energy beamforming gain versus the number of feedback intervals $N$ under energy measurement errors.}
\label{fig:error2}
\end{figure}

Fig. \ref{fig:error1} shows the normalized error of estimated matrix norm versus the number of feedback intervals $N$ subject to different values of energy measurement errors with $B=2$. It is observed that for both channel learning schemes, the measurement errors with $\alpha = 0.001$ and $\alpha = 0.01$ result in error floors for channel estimation. In addition, when $N \ge 60$, energy comparison is observed to perform worse than energy quantization in the case with $\alpha = 0.01$, which is in sharp contrast to the case without energy measurement errors (i.e., $\alpha = 0$), where the reverse is true. This shows that energy comparison is generally more sensitive to the energy measurement errors than energy quantization.

Fig. \ref{fig:error2} shows the resulting energy beamforming gain versus $N$ under different levels of energy measurement errors with $B=2$. It is observed that with $\alpha = 0.001$ and $\alpha = 0.01$, the two channel learning schemes have similar energy beamforming gains as that without energy measurement errors (i.e., $\alpha = 0$), despite the error floor phenomenon observed for the normalized channel matrix in Fig. \ref{fig:error1}. This shows that in MIMO WET systems, our proposed channel learning schemes (with above robust design) are immune to the energy measurement errors.

\section{Conclusion}\label{sec:conclusion}

This paper studied a new form of energy measurement based feedback for channel estimation in MIMO WET systems, by taking into account the energy and hardware limitations of practical ERs. We proposed a general channel learning design framework, by leveraging the technique of ACCPM in convex optimization. Under this framework, we developed two energy feedback schemes, namely energy quantization and energy comparison, for efficiently encoding the energy measurements at the ER with limited feedback. Simulation results showed that energy quantization is generally more effective when the number of feedback bits per interval is large, while energy comparison is preferable with small number of feedback bits. It is our hope that this paper provides new insights and will inspire future investigations on the practical design of channel learning for MIMO WET systems or other wireless communication systems based on the signal energy/power feedback.


\begin{thebibliography}{1}
\bibliographystyle{IEEEbib}

\bibitem{BiHoZhang2014}
S. Bi, C. K. Ho, and R. Zhang, ``Wireless powered communication: opportunities and challenges,'' {\it IEEE Commun. Mag.}, vol. 53, no. 4, pp.117-125, Apr. 2015.

\bibitem{BiZengZhang2016}
S. Bi, Y. Zeng, and R. Zhang, ``Wireless powered communication networks: an overview,'' {\it IEEE Wireless Commun.}. [Online] Available: {\url{http://arxiv.org/abs/1508.06366}}

\bibitem{ZhouZhangHo2013}
X. Zhou, R. Zhang, and C. K. Ho, ``Wireless information and power transfer: architecture design and rate-energy tradeoff,'' {\it IEEE Trans. Commun.}, vol. 61, no. 11, pp. 4757-4767, Nov. 2013.

\bibitem{Powercast}
Powercast, \url{http://www.powercastco.com/}

\bibitem{ZhangHo2013}
R. Zhang and C. K. Ho, ``MIMO broadcasting for simultaneous wireless information and power transfer,'' {\it IEEE Trans. Wireless Commun.}, vol. 12, no. 5, pp. 1989-2001, May 2013.

\bibitem{XuLiuZhang2014}
J. Xu, L. Liu, and R. Zhang, ``Multiuser MISO beamforming for simultaneous wireless information and power transfer,'' {\it IEEE Trans. Signal Process.}, vol. 62, no. 18, pp. 4798-4810, Sep. 2014.

\bibitem{JuZhang2014}
H. Ju and R. Zhang, ``Throughput maximization in wireless powered communication networks,'' {\it IEEE Trans. Wireless Commun.}, vol. 13, no. 1, pp. 418-428, Jan. 2014.



\bibitem{LiuZhangChua2014}
L. Liu, R. Zhang, and K. C. Chua, ``Multi-antenna wireless powered communication with energy beamforming,'' {\it IEEE Trans. Commun.}, vol. 62, no. 12, pp. 4349-4361, Dec. 2014.

\bibitem{XuBiZhang2015}
J. Xu, S. Bi, and R. Zhang, ``Multiuser MIMO wireless energy transfer with coexisting opportunistic communication,'' {\it IEEE Wireless Commun. Letters}, vol. 4, no. 3, pp. 273-276, Jun. 2015.

\bibitem{CheXuDuanZhang2015}
Y. Che, J. Xu, L. Duan, and R. Zhang, ``Multiantenna wireless powered communication with co-channel energy and information transfer,'' {\it IEEE Commun. Letters}, vol. 19, no. 12, pp. 2266-2269, Dec. 2015.

\bibitem{ZengZhang2014A}
Y. Zeng and R. Zhang, ``Optimized training design for wireless energy transfer,'' {\it IEEE Trans. Commun.}, vol. 63, no. 2, pp. 536-550, Feb. 2015.

\bibitem{ZengZhang2014B}
Y. Zeng and R. Zhang, ``Optimized training for net energy maximization in multi-antenna wireless energy transfer over frequency-selective channel,'' {\it IEEE Trans. Commun.}, vol. 63, no. 6, pp. 2360-2373, Jun. 2015.

\bibitem{YangHoZhangGuan2014}
G. Yang, C. K. Ho, R. Zhang, and Y. L. Guan, ``Throughput optimization for massive MIMO systems powered by wireless energy transfer,'' {\it IEEE J. Sel. Areas Commun.}, vol. 33, no. 8, pp. 1640-1650, Aug. 2015.

\bibitem{YangHoGuan2014}
G. Yang, C. K. Ho, and Y. L. Guan, ``Dynamic resource allocation for multiple-antenna wireless power transfer,'' vol. 62, no. 14, pp. 3565-3577, Jul. 2014.

\bibitem{Love2008}
D. J. Love, R. W. Heath Jr., V. K. N. Lau, D. Gesbert, B. D. Rao, and M. Andrews, ``An overview of limited feedback in wireless communication systems,'' {\it IEEE J. Sel. Areas Commun.}, vol. 26, no. 8, pp. 1341-1365, Oct. 2008.

\bibitem{XuZhang_OneBit}
J. Xu and R. Zhang, ``Energy beamforming with one-bit feedback,'' {\it IEEE Trans. Signal Process.}, vol. 62, no. 20, pp. 5370-5381, Oct. 2014.

\bibitem{NoamGoldsmith2013}
Y. Noam and A. Goldsmith, ``The one-bit null space learning algorithm and its convergence,'' {\it IEEE Trans. Signal Process.}, vol. 61, no. 24, pp. 6135-6149, Dec. 2013.

\bibitem{GopalakrishnanSidiropoulos2014}
B. Gopalakrishnan and N. D. Sidiropoulos, ``Cognitive transmit beamforming from binary CSIT,'' {\it IEEE Trans. Wireless Commun.}, vol. 14, no. 2, pp. 895-906, Feb. 2015.

\bibitem{ACCPM}
S. Boyd, ``Convex optimization II,'' Stanford University. [Online]. Available: {\url{http://www.stanford.edu/class/ee364b/lectures.html}}

\bibitem{BV:Convex}
S. Boyd and L. Vandenberghe, {\it Convex Optimization}. Cambridge, U.K.: Cambridge Univ. Press, 2004.

\bibitem{cvx}
M. Grant and S. Boyd, {\it CVX: Matlab software for disciplined convex programming, version 1.21}, {\url{http://cvxr.com/cvx/}}, Apr. 2011.


\bibitem{Quantization}
R. M. Gray and D. L. Neuhoff, ``Quantization,'' {\it IEEE Trans. Inform. Theory}, vol. 44, no. 6, pp. 2325-2383, Oct. 1998.
\end{thebibliography}
\end{document}